\documentclass[12pt, preprint]{aastex}
\usepackage{epsfig}
\usepackage{natbib}

\newcommand{\Hm}{\rm{H}^{-}}
\newcommand{\Hp}{\rm{H}^{+}}

\newcommand{\mHtp}{\rm{H}_{2}^{+}}
\newcommand{\htp}{\rm{H}_{3}^{+}}
\newcommand{\mH}{\rm{H}}
\newcommand{\mHt}{\rm{H}_{2}}

\newcommand{\xhteq}{x_{\rm H_{2}, eq}}
\newcommand{\Htvol}{\langle x_{\rm{H}_{2}}\rangle_{\rm V}}
\newcommand{\Htmass}{\langle x_{\rm{H}_{2}}\rangle_{\rm M}}

\newcommand{\msun}{{\rm M_{\odot}}}
\def\simless{\mathbin{\lower 3pt\hbox
   {$\rlap{\raise 5pt\hbox{$\char'074$}}\mathchar"7218$}}}   
\def\simgreat{\mathbin{\lower 3pt\hbox  
   {$\rlap{\raise 5pt\hbox{$\char'076$}}\mathchar"7218$}}} 

\begin{document}

\title{Simulating the formation of molecular clouds. II.
Rapid formation from turbulent initial conditions}

\author{Simon~C.~O. Glover$^{1,2}$ \& Mordecai-Mark {Mac Low}$^1$}
\affil{$^1$Department of Astrophysics, American Museum of Natural History, \\
       Central Park West at 79th Street, New York, NY 10024}
\affil{$^2$Astrophysikalisches Institut Potsdam,\\ An der Sternwarte 16, D-14482
  Potsdam, Germany}
\email{sglover@aip.de, mordecai@amnh.org}

\begin{abstract}
The characteristic lifetimes of molecular clouds remain uncertain and
subject to debate, with arguments having recently been advanced in 
support of short-lived clouds, with lifetimes of only a few Myr, and in 
support of much longer-lived clouds, with lifetimes of 10~Myr or more. 
One argument that has been advanced in favour of long cloud lifetimes is the
apparent difficulty involved in converting sufficient atomic hydrogen to molecular 
hydrogen within the short timescale required by the rapid cloud formation scenario. 
However, previous estimates of the time required for this conversion to occur have 
not taken into account the effects of the supersonic turbulence which is inferred to 
be present in the atomic gas. 

In this paper, we present results from a large set of numerical
simulations that demonstrate that $\mHt$ formation occurs rapidly in
turbulent gas.  Starting with purely atomic hydrogen, large quantities
of molecular hydrogen can be produced on timescales of 1--2~Myr, given
turbulent velocity dispersions and magnetic field strengths consistent
with observations. Moreover, as our simulations underestimate the
effectiveness of H$_2$ self-shielding and dust absorption, we can be
confident that the molecular fractions that we compute are strong
lower limits on the true values. The formation of large quantities of
molecular gas on the timescale required by rapid cloud formation
models therefore appears to be entirely plausible.

We also investigate the density and temperature distributions of gas
in our model clouds. We show that the density probability distribution
function is approximately log-normal, with a dispersion that agrees
well with the prediction of Padoan, Nordlund \& Jones~(1997). The
temperature distribution is similar to that of a polytrope, with an
effective polytropic index $\gamma_{\rm eff} \simeq 0.8$, although at
low gas densities, the scatter of the actual gas temperature around
this mean value is considerable, and the polytropic approximation does
not capture the full range of behaviour of the gas.
\end{abstract}


\section{Introduction}
\label{intro}
An important goal of star formation research is the development of an understanding
of the formation of molecular clouds, since these clouds host all observed Galactic 
star formation. The main constituent of these clouds is molecular hydrogen, $\mHt$,
which forms in the interstellar medium (ISM) on the surface of dust grains, as its 
formation in the gas phase by radiative association occurs at a negligibly slow rate
\citep{gs63}. The basic physics of $\mHt$ formation on dust is believed to be well
understood, although uncertainties remain about various issues such as the relative
importance of physisorption versus chemisorption \citep[see e.g.][]{katz99,ct04} or the 
role of dust grain porosity \citep{pb06}.

However, many other questions remain unanswered. In particular, there is little 
consensus regarding the physical mechanism responsible for cloud formation. 
One school of thought holds that molecular clouds are transient objects, formed and
then dispersed on a timescale of only a few million years by the action
of large-scale turbulent flows in the interstellar medium (ISM), which are themselves
driven by energy input from supernovae and from the magnetorotational instability
\citep{mk04}. Several lines of evidence in favour of this picture, such as the 
absence of post-T Tauri stars with ages greater than $3 \: {\rm Myr}$ in local star 
forming regions, are discussed by \citet{hbpb01}. 

An alternative school of thought holds that clouds are gravitationally bound objects
in virial equilibrium which evolve on timescales of the order of $10 \: {\rm Myr}$ or
longer,
and are supported by magnetic pressure \citep{tm04,mtk06}, or by
turbulence driven by the expansion of internal H~{\sc ii} regions\citep{m02,kmm06}. 
In this picture, it is easier to reconcile the observed
mass of molecular gas in the Galaxy ($M_{\mHt} \sim 10^{9} \: \msun$; \citealt{evans99}) 
with the inferred Galactic star formation rate ($\dot{M}_{*} \sim 3 \: \msun \: {\rm yr^{-1}}$; 
\citealt{scalo86}) without requiring the average star formation efficiency of molecular 
clouds to be very small. 

One argument often advanced in favour of longer cloud lifetimes is the apparent 
difficulty involved in producing sufficient $\mHt$ in only one or two Myr to explain
observed clouds, given the relatively slow rate at which $\mHt$ forms in the ISM.
The $\mHt$ formation timescale in the ISM is approximately \citep{hws71}
\begin{equation}
 t_{\rm form} \simeq \frac{10^{9} \: {\rm yr}}{n},
\end{equation}
where $n$ is the number density in ${\rm cm^{-3}}$, which suggests that in gas
with a mean number density $\bar{n} \sim 100 \: {\rm cm^{-3}}$, characteristic of 
most giant molecular clouds \citep{bs80}, conversion from atomic to molecular form 
should take at least $10 \: {\rm Myr}$, longer than the entire lifetime of a transient 
cloud. However, estimates of this kind do not take account of dynamical processes such 
as supersonic turbulence or thermal instability which may exert a great deal of 
influence on the effective $\mHt$ formation rate. The impact of these processes can 
be investigated through the use of numerical modelling, but in most models published 
to date either the dynamics of the gas have been highly simplified, with the flow 
restricted to one or two dimensions and with only highly ordered large-scale flows 
considered \citep[see e.g.][]{hp99,hp00,ki00,ki02,berg04}, or insufficient chemistry 
has been included to address the question at hand (e.g.\ \citealt{kor99}; \citealt{da00};
\citealt{wada01}; \citeauthor{kn02a}~2002a,b,~2004; \citealt{bkmm04};
\citealt{dab04};  \citealt{sdbs05}; \citealt{jm06}).

In \citet[][hereafter paper I]{gm06} we described how we have modified the publicly 
available ZEUS-MP hydrodynamical code \citep{norman00} to allow us to accurately
simulate the thermal balance of the atomic and molecular gas in the ISM, and to
follow the growth of the molecular hydrogen fraction with time. An important feature
of our approach is the use of simple approximations to treat the effects of $\mHt$ 
self-shielding and dust shielding, allowing us to approximate the $\mHt$ destruction 
rate without requiring us to solve the radiative transfer equation for the 
photodissociating radiation, which could only be done at a prohibitively high 
computational cost. 

We also presented results from some simple tests of our modified code and showed
that it produces physically reasonable results. Finally, we used our modified code
to model molecular cloud formation by the gravitational collapse of quasi-uniform,
initially static gas. We showed that in this scenario, large-scale conversion of atomic
to molecular gas occurs on a timescale $t > 10 \: {\rm Myr}$, equivalent to at least
1--2 free-fall times (for simulations performed with our canonical value for the initial 
gas number density, $n_{\rm i} = 100 \: {\rm cm^{-3}}$).

In the current paper, we present results from a set of simulations performed with a
supersonically turbulent initial velocity field, and show that the presence of turbulence
dramatically reduces the time required to form large quantities of $\mHt$. In
supersonically turbulent gas, the large density compressions created by the turbulence
allow $\mHt$ to form rapidly, with large molecular fractions being produced after only
1--2~Myr, consistent with the timescale required by rapid cloud formation models. 
We also find evidence that much of the $\mHt$ is formed in high density gas and then
transported to lower densities by the action of the turbulence, a phenomenon that
may have a significant impact on the chemistry of the ISM \citep{xal95,wla02}. 

The plan of this paper is as follows. In \S~\ref{numeric}, we briefly describe the
numerical method we use to simulate the gas and in \S~\ref{init_cond} we 
discuss the initial conditions used for our simulations. In \S~\ref{timescale},
we examine the timescale for $\mHt$ formation in turbulent gas, and in 
\S~\ref{mol_distrib} we analyze the resulting distribution of the molecular gas.
In \S~\ref{thermo}, we discuss the temperature distribution of the gas and its
evolution with time, and also examine the behaviour of the effective polytropic index,
$\gamma_{\rm eff}$. In \S~\ref{sense} we examine how sensitive our results
are to variations of our initial conditions. We conclude in \S~\ref{summary}
with a summary of our results.

\section{Numerical method}
\label{numeric}
We solve the equations of fluid flow for the gas in the ISM using a modified version of
the ZEUS-MP hydrodynamical code \citep{norman00}.  Our modifications are based on
those developed for ZEUS-3D by \citet{pav02} and \citet{sr03}, which in turn are built on
earlier work by \citet{sutt97}, and involve two major changes to the code.

First, we have added a limited treatment of non-equilibrium chemistry to the code. 
We treat the chemistry in an operator split fashion. During the source step, we solve the
coupled set of chemical rate equations for the fluid, together with the portion of the
internal energy equation dealing with compressional and radiative heating and cooling,
under the assumption that the other hydrodynamical variables (e.g.\ density) remain
fixed. During the subsequent advection step, we advect tracer fields which track the 
abundances of the chemical species of interest and which advect as densities, while
the internal energy is advected as it would be in the unmodified version of ZEUS-MP.
For stability, we solve the coupled equations implicitly, using Gauss-Seidel iteration.
On the rare occasions that the iteration fails to converge, we use a more expensive
bisection algorithm that is guaranteed to find a solution. 

To improve the efficiency of the code, we make use of subcycling. Since the cooling
time is generally much shorter than the chemical timescale for gas with the range of
physical conditions of interest in this study, we evolve both the chemical rate equations
and the energy equation on a cooling timestep, $\Delta t_{\rm cool}$, given by:
\begin{equation}
\Delta t_{\rm cool} = 0.3 \frac{e}{|\Lambda|},
\end{equation}
where $\Lambda$ is the net rate at which the gas gains or loses energy due to radiative
and chemical heating and cooling: in the convention used in our code, $\Lambda < 0$ 
corresponds to cooling and $\Lambda > 0$ to heating. Often, $\Delta t_{\rm cool}$ will
be much shorter than the hydrodynamical timestep, $\Delta t_{\rm hydro}$, which is 
calculated as detailed in \citet{sn92a}. In that case, rather than constraining the code to
evolve the hydrodynamics on the cooling timestep, we allow the chemistry code to
subcycle: to take multiple source steps, each of length $\Delta t_{\rm cool}$, until the
total elapsed time is equal to $\Delta t_{\rm hydro}$. The cooling time is recomputed
following every chemistry and cooling substep, and the code also ensures that the
total elapsed time cannot exceed $\Delta t_{\rm hydro}$ by shortening the last substep
(if required). To further improve the efficiency of the code, our implementation of subcycling
functions at the level of individual grid zones, so that only zones for which 
 $\Delta t_{\rm cool} < \Delta t_{\rm hydro}$ are subcycled.

In our chemical model, we adopt standard solar abundances for hydrogen and helium, 
together with metal abundances taken from \citet{sem00}. We assume that metals with 
ionization potential less than that of atomic hydrogen (e.g.\ carbon, silicon) remain singly 
ionized throughout the calculation,  and that metals with ionization potentials greater than
that of hydrogen (e.g.\ helium, nitrogen) remain neutral. The ionization state of oxygen, which has 
very nearly the same ionization potential as hydrogen, is assumed to track that of hydrogen, 
owing to the influence of rapid charge transfer reactions.
We also assume that minor ions such as $\Hm$, $\mHtp$ or $\htp$ can be ignored, and 
so are left with only four species of interest: free electrons, $\Hp$, $\mH$ and $\mHt$.
The abundances of these four species are constrained by two conservation laws --
conservation of charge and of the number of hydrogen nuclei -- and so only two of these
species need to be tracked directly. We choose to follow the abundances of $\mHt$ and
$\Hp$. The reactions included in our chemical model, and further
details including justification of our simplifying approximations, are given in
Paper I. We discuss photodissociation further below.

The second major modification to the ZEUS-MP code that we have made is to incorporate
a detailed treatment of the radiative and chemical heating and cooling
of the gas. 
%
The main heat source in gas with a low level of extinction is photoelectric heating (caused 
by the ejection of photoelectrons from dust grains and polycyclic aromatic hydrocarbons
illuminated by photons with energies in the range $6 < h\nu < 13.6 \: {\rm eV}$), while in 
gas with high extinction, cosmic ray heating dominates. Our treatment of these processes 
is based on \citet{bt94} (as modified by \citealt{w03}) and \citet{gl78} respectively, with 
the effects of extinction modelled using the approximation discussed in \S~\ref{shield} below.

The main coolants at the temperatures and densities of interest in our simulations are the 
fine structure lines of ${\rm C^{+}}$,  ${\rm O}$ and ${\rm Si^{+}}$. We compute the cooling 
from these species exactly, using atomic data from a variety of
sources (see Paper I). 
We also include heating and cooling from a variety of other 
processes, which are generally of lesser importance, and which are also summarized in 
Paper I.

\subsection{${\mathbf H_{2}}$ photodissociation, self-shielding and dust extinction}
\label{shield}
Following \citet{db96}, we assume that the $\mHt$ photodissociation rate can be
written as the sum of an optically thin rate, $k_{\rm ph, 0}$, and two separate 
shielding factors, $f_{\rm shield}$ and $f_{\rm dust}$ that account for the effects of
$\mHt$ self-shielding and dust shielding respectively:
\begin{equation}
 k_{\rm ph} = k_{\rm ph, 0} f_{\rm shield} f_{\rm dust}.
\end{equation}
For the optically thin rate, we adopt the expression
\begin{equation}
 k_{\rm ph, 0} = 3.3 \times 10^{-11} \chi \: {\rm s}^{-1}, \label{kph_thin}
\end{equation}
where we have assumed that the ultraviolet (UV) field has the same spectral shape
as the \citet{d78} field, and where $\chi$ is a dimensionless factor which
characterizes the intensity of the field at $1000\:$\AA~relative to the 
\citet{habing68} field; note that for the original \citeauthor{d78} field, $\chi = 1.7$.

To treat the effects of self-shielding, we use the approximation for $f_{\rm shield}$
suggested by \citet{db96}:
\begin{equation}
f_{\rm shield} = \frac{0.965}{(1 + x/b_{5})^{2}} + \frac{0.035}{(1 + x)^{1/2}}
 \exp\left[-8.5 \times 10^{-4} (1+x)^{1/2}\right], \label{fsh}
\end{equation}
where $x = N_{\mHt} / 5 \times 10^{14} \: {\rm cm}^{-2}$, with $N_{\mHt}$ 
being the $\mHt$ column density between the fluid element and the source
of the UV radiation, and with $b_{5} = b / 10^{5} \: {\rm cm}
\:{\rm s}^{-1}$, where $b$ is the Doppler broadening parameter.  

As in paper I, we use two different approximation to compute $N_{\mHt}$
for every grid zone. In most of our simulations, we use a local
shielding approximation. In this approximation, we assume that the
dominant contribution to the self-shielding of any given fluid element
comes from gas in the immediate vicinity of that element, and
consequently approximate $N_{\mHt}$ as:
\begin{equation}
 \tilde{N}_{\mHt} = \frac{\Delta x}{2} n_{\mHt},
\end{equation}
where $n_{\mHt}$ is the $\mHt$ number density in the zone of interest and 
$\Delta x$ is the width of the zone, measured parallel to one of the coordinate
axes. Note that since our grid zones are cubical, the choice of axis is immaterial.
This approximation is intended to capture, in a crude fashion, the effects on $\mHt$
self-shielding of the significant Doppler shifts along any particular
line of sight. 

In a few simulations, however, we use an approach that we referred
to in paper I as the six-ray shielding approximation. In this
approach, we compute an effective $\mHt$ column density for each
grid zone by first computing exact values along a small number of lines 
of sight and then averaging these values appropriately. To simplify
the implementation, the lines of sight are chosen to be parallel to 
the coordinate axes of the grid. An approach of this type has previously 
been used by \citet{nl97,nl99} and \citet{yahs03}.

To treat shielding due to dust, we use a similar approach. We follow
\citet{db96} and write $f_{\rm dust}$ as 
\begin{equation}
f_{\rm dust} = e^{-\tau_{\rm d, 1000}} = e^{-\sigma_{\rm d, 1000} N_{\rm \mH, tot}},
\end{equation}
where $\tau_{\rm d, 1000}$ is the optical depth due to dust at a wavelength of $1000 $\AA, 
$\sigma_{\rm d, 1000}$ is the corresponding effective attenuation cross-section 
(which has a value $\sigma_{\rm d, 1000} \simeq 2 \times 10^{-21} \: {\rm cm^{2}}$
for the diffuse ISM), and $N_{\rm \mH, tot}$ is the total column density of hydrogen
nuclei between the zone of interest and the source of the UV. To compute 
$N_{\rm \mH, tot}$, we use the same approximations as described above. In other 
words, when we use the local shielding approximation, we assume that the dominant
contribution to $N_{\rm \mH, tot}$ comes from local gas, and so write it as:
\begin{equation}
\tilde{N}_{\rm \mH, tot} = \frac{\Delta x}{2} \left( n_{\mH} + n_{\Hp} + 
2 n_{\mHt} \right),
\end{equation}
while in the six-ray approximation, we again compute it exactly along a small
number of lines of sight.

The main advantage of our local shielding approximation (aside from its
simplicity and speed) is the fact that it allows us to be certain that 
we are {\em underestimating} the true amount of shielding and hence 
{\em overestimating} the $\mHt$ photodissociation rate in simulations
that use it.  We can therefore be confident that the $\mHt$ fractions 
computed in these simulation are lower limits on the true values and hence that 
the timescales for $\mHt$ formation that we find are {\em upper limits} on the 
actual values. As we will see later, this fact serves to strengthen the 
conclusions that we draw from these simulations.

A major disadvantage of our local shielding approximation is that they
make the photodissociation rate depend explicitly on the numerical resolution 
$\Delta x$. Consequently, the equilibrium $\mHt$ abundance, $\xhteq$, also becomes 
resolution dependent, as can easily be seen from the following equation for $\xhteq$
\begin{equation}
 \frac{\xhteq}{1 - \xhteq} = \frac{2 R_{\rm form}}{R_{\rm ph}} n, \label{eq:xhteq}
\end{equation}
where $n$ is the number density of hydrogen nuclei, $R_{\rm form}$ is the formation 
rate of $\mHt$ on dust grain surfaces, $R_{\rm ph} = k_{\rm ph} n_{\mHt}$, and where 
the $\mHt$ abundance $x_{\mHt}$ is defined such that $x_{\mHt} = 2 n_{\mHt} / n$.
It must also be said that the physical justification for treating dust shielding
using the local approximation is somewhat lacking since the shielding provided
by the dust should not be significantly affected by Doppler shifts along the line
of sight. In paper I, we saw that for initially static, uniform density gas, 
these drawbacks are very serious and results obtained using the local 
approximation are of questionable accuracy. However, in the turbulent flows 
studied in the present paper, the performance of the local shielding 
approximation is very much better, as we will see in \S~\ref{timescale} \&
\ref{mol_distrib} below.

A major disadvantage of the six-ray approximation is the extremely coarse 
angular resolution of the radiation field that it provides. This poor
angular resolution will cause us to overestimate the amount of 
shielding in some regions, and underestimate it in others: the
precise details will depend on the particular form of the density 
field, but in general we will tend to underestimate the amount of
shielding whenever the volume filling factor of dense gas is small.
On the other hand, the fact that this approach does not take account of 
velocity structure along any of the lines of sight means that it may
significantly overestimate $f_{\rm shield}$ in a supersonically
turbulent flow. It is therefore difficult to determine whether the
amount of $\mHt$ produced in simulations using this method is an
overestimate or an underestimate of the true amount. For the main 
problem that we are interested in investigating -- the determination 
of the $\mHt$ formation rate in dynamically evolving, cold atomic gas 
-- this is problematic, as it may lead us to derive an artificially 
short timescale for $\mHt$ formation. In view of this, we have adopted
the local shielding approximation in most of our simulations, and have
run only a few simulations using the six-ray approximation for the
purposes of comparison. As we will see later, the results of these
simulations agree surprisingly well with those of simulations using
the local approximation.

Finally, we note that to compute the effect of the dust shielding on the
photoelectric heating rate, we again use a very similar approximation: we
use our value of $N_{\rm \mH, tot}$ computed above to calculate $A_{V}$,
the extinction of the gas in the V band, and then compute the photoelectric
heating rate using a radiation field strength attenuated by a factor 
$e^{-2.5A_{\rm V}}$, as suggested by \citet{berg04}.

\section{Initial conditions}
\label{init_cond}

\subsection{Box size and initial number density} 
Since our aim in this paper is to model the transition from atomic to 
molecular gas, we have chosen to consider relatively small volumes,
which we visualize as being smaller sub-regions within larger, gravitationally 
collapsing clouds of gas, such as those found in the simulations of \citet{krav03}
or \citet{lmk05}. We therefore consider a cubical volume of size $L$, and
apply periodic boundary conditions to all sides of the cube. As we discussed
in paper I, in choosing a value for $L$ we aimed to strike a reasonable balance
between simulating a large, representative region of a cloud (which argues for
large $L$) and maximizing our physical resolution for any given numerical
resolution (which argues for small $L$). For most of the simulations presented
here, we settled on $L = 20 \: {\rm pc}$ as an appropriate value, but in 
\S~\ref{turb_box} we examine the effects of varying $L$. The value of $L$ used in 
each simulation is summarized in Table~\ref{turb_runs}, together with the values used for
our other adjustable parameters.

Within the box, we assumed an initially uniform distribution of gas, 
characterized by an initial number density $n_{\rm i}$. In most of our simulations,
we take $n_{\rm i} = 100 \: {\rm cm^{-3}}$ as this value is consistent with the 
inferred mean densities of many observed molecular clouds \citep{mk04}.
However, in \S~\ref{turb_n0} we explore the effects of reducing $n_{\rm i}$. 
Atomic gas with $n = 100 \: {\rm cm^{-3}}$ lies well within the cold neutral medium
regime \citep{w95,w03} and has a short cooling time ($t_{\rm cool} < 0.05 
\: {\rm Myr}$), and so our results are insensitive to the initial temperature
$T_{\rm i}$. In most of our simulations, we adopt a rather arbitrary initial 
temperature $T_{\rm i} = 1000 \: {\rm K}$, but as we demonstrate in 
\S~\ref{temp}, simulations with $T_{\rm i} = 100 \: {\rm K}$ produce 
essentially  identical results for times $t \simgreat  0.05 \: {\rm Myr}$.

\subsection{Magnetic field strength}
Since there is now considerable observational evidence for the presence of 
dynamically significant magnetic fields in interstellar gas \citep{beck01,hc05},
we included a magnetic field in the majority of our simulations. For simplicity, 
in simulations where a field was present, we assumed that it was initially 
uniform and oriented parallel to the $z$-axis of the simulation. The strength of 
the field was a free parameter, and the values used in our various simulations 
are summarized in Table~\ref{turb_runs}. Observational determinations of the 
local magnetic field strength give a typical value of $6 \pm 2 \mu$G, and so we 
ensured that our fiducial value for the initial magnetic field strength, 
$B_{i, {\rm fid}} = 5.85 \mu{\rm G}$, was consistent with this value.
 
Our choice of this slightly odd value was motivated by our desire that the 
mass-to-flux ratio of the gas should be some simple multiple of the critical
value \citep{nn78}
\begin{equation}
 \left( \frac{M}{\Phi} \right)_{\rm crit} = \frac{1}{2\pi \sqrt{G}},
\end{equation}
at which magnetic pressure balances gravity in an isothermal slab. 
For a fully atomic cloud of gas, the mass-to-flux ratio can be written in units
of this critical value as \citep{cnwk04}
\begin{equation}
 \frac{M}{\Phi} = 3.8 \times 10^{-21} \frac{N_{\mH}}{B}, \label{m2f-atom}
\end{equation}
where $N_{\mH}$ is column density of atomic hydrogen in units of ${\rm cm^{-2}}$ 
and $B$ is the strength of the magnetic field, in units of $\mu{\rm G}$. For a 
simulation with a box size of $20 \: {\rm pc}$ and an initial atomic hydrogen 
number density $n_{\rm i} = 100 \: {\rm cm^{-3}}$, this gives $M / \Phi = 23.45 / B_{i}$, 
where $B_{i}$ is the initial magnetic field strength (in units of $\mu{\rm G}$), 
and so if $B_{i} = B_{i, {\rm fid}}$, then $(M / \Phi)_{\rm fid} = 4$.
Observations of magnetic field strengths in molecular cloud cores, summarized in 
\citet{cru99} and \citet{cnwk04}, find a smaller mean value, $(M / \Phi)_{\rm mean} 
\simeq 2$, but there is significant scatter around this value, and in any case
it is far from clear that we would expect to find the same value of $M / \Phi$
in dense cloud cores as we would find in the much lower density neutral atomic 
gas. 

In addition to this fiducial case, we ran simulations with a number of other initial
field strengths, as detailed in Table~\ref{turb_runs} and in \S~\ref{mag_field}.

\subsection{Initial velocity field}
We generate the initial velocity field required for our simulations of turbulent gas by
using the method described in \citet{mkbs98} and \citet{mac99}. In this method, we 
produce values for each component of velocity in each grid zone by drawing values from
a Gaussian random field constructed to have a flat power spectrum for wavenumbers 
$k \equiv L / \lambda_{\rm d}$ in the range $1 \leq k \leq 2$ and zero power outside of 
that range. The normalization of the resulting velocity field was then adjusted until the 
root mean square velocity, $v_{\rm rms}$, matched a previously specified initial value, 
$v_{\rm rms, i}$.

In most of our simulations, we set $v_{\rm rms, i} = 10 \: {\rm km} \: {\rm s^{-1}}$. 
Our choice of this value is motivated by observations of molecular clouds on scales 
$\sim 20 \: {\rm pc}$ \citep{srby87} which find line-of-sight velocity dispersions 
$\sigma_{\rm los} \simeq 2$--$7 \: {\rm km} \: {\rm s^{-1}}$. If due primarily to turbulent 
motions in the clouds, these velocity dispersions imply RMS turbulent velocities in the 
range $v_{\rm rms} \simeq 3.5$--$12 \: {\rm km} \: {\rm s^{-1}}$. Our chosen value lies 
towards the top end of this range, but in \S~\ref{turb_vrms} we examine the effects 
of adopting smaller initial values.

Finally, it should be noted that in the simulations presented in this paper, we consider 
only decaying turbulence, i.e.\  turbulence which is not maintained by a regular input of 
kinetic energy, and which therefore largely dissipates within a few turbulent crossing 
times \citep{mac99}. However, we find that significant amounts of $\mHt$ form within only
1--$2 \: {\rm Myr}$, before the RMS velocity has decayed by more than a factor of two.
We therefore anticipate that the results for driven turbulence would be quite similar.
We hope to investigate this further in future work.

\section{The ${\mathbf H_{2}}$ formation timescale}
\label{timescale}
As in paper I, we first consider the formation of $\mHt$ in some detail in a set of 
fiducial simulations differing only in numerical resolution, before going on to 
consider in later sections the effects of changing the various input 
parameters. Parameters for the full set of simulations are listed in Table~\ref{turb_runs}.
The values we adopted for our fiducial runs are an initial density 
$n_{\rm i} = 100 \: {\rm cm^{-3}}$, an initial temperature $T_{\rm i} = 1000 \: {\rm K}$,
an initial RMS turbulent velocity $v_{\rm rms, i} = 10 \: {\rm km} \: {\rm s^{-1}}$,
an initial magnetic field strength $B_{\rm i} = 5.85 \mu{\rm G}$ and a box size
$L = 20 \: {\rm pc}$. With these parameters, there is initially $\sim 0.1 \: M_{\rm J}$
of gas in the box, but this number rapidly increases to $\sim 10 \: M_{\rm J}$
as the gas cools to a thermal equilibrium temperature of approximately $65 \: {\rm K}$. 
This initial period of cooling takes place in less than 0.05~Myr, just as in the 
initially static simulations discussed in paper I.
This is a much shorter timescale than the turbulent crossing time of the box,
$t_{\rm cross} = L / v_{\rm rms, i} \simeq 2 \: {\rm Myr}$, and so the gas reaches 
thermal equilibrium before the turbulence has had time to strongly perturb the 
density structure. The initial mass-to-flux ratio of the gas in the box is 
$M / \Phi = 4$, in units of the critical value. 

\begin{figure}
\centering
\epsfig{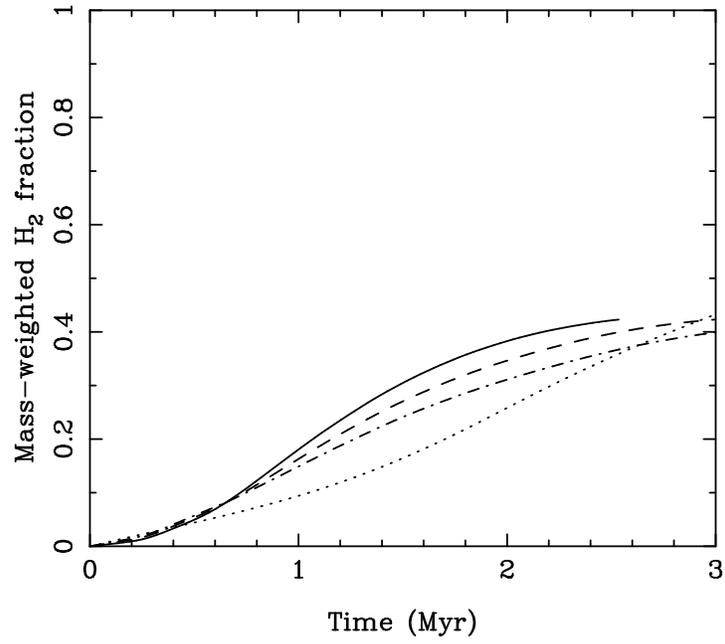}
\caption{Evolution of $\Htmass$ with time in four runs with increasing 
numerical resolution: run MT64 (dotted line), run MT128 (dot-dashed line), 
run MT256 (dashed line) and run MT512 (solid line).}
\label{H2-MT-fid}
\end{figure}

We performed four runs with these parameters, with numerical resolutions of 
$64^{3}$, $128^{3}$, $256^{3}$ and $512^{3}$ zones, which we designate as
MT64, MT128, MT256 and MT512 respectively. In Figure~\ref{H2-MT-fid}, we
plot the evolution with time in these four runs of the mass-weighted mean 
molecular fraction, $\Htmass$, defined as
\begin{equation}
\Htmass = \frac{\sum_{i,j,k} \rho_{\mHt}(i,j,k) \Delta V(i,j,k)}{M_{\rm H}},
\end{equation}
where $\rho_{\mHt}(i,j,k)$ is the mass density of $\mHt$ in zone $(i,j,k)$,
$\Delta V(i,j,k)$ is the volume of zone $(i,j,k)$,
$M_{\rm H}$ is the total mass of hydrogen present in the simulation, and
where we sum over all grid zones.

As Figure~\ref{H2-MT-fid} makes clear, the $\mHt$ abundance evolves rapidly in 
turbulent gas. Large quantities of $\mHt$ are produced within a short time, with 
mass-weighted mean molecular fractions of approximately 40\% being produced within 
only $2 \: {\rm Myr}$. For comparison, in our fiducial static runs in paper I,
we found that $\Htmass$ exceeded 40\% only for $t > 30.8 \: {\rm Myr}$ in the run
performed using the local shielding approximation (run MS256 in the notation of
paper I), and at $t > 7.6 \: {\rm Myr}$ in the run performed using the
six-ray shielding approximation (run MS256-RT). The latter time is 
probably the better estimate, but even in this case there is a factor of 3--4
difference in the $\mHt$ formation timescale between the static and turbulent
runs. Indeed, $\mHt$ forms significantly faster in our fiducial turbulent runs
than in a static run in which there was no ultraviolet background and hence
no photodissociation of $\mHt$ (run MS256-nr; in this run, the time at which
$\Htmass > 0.4$ is again approximately $7.6 \: {\rm Myr}$). It is also apparent 
from Figure~\ref{H2-MT-fid}
that although the evolution of $\Htmass$ in these simulations shows some 
dependence on the numerical resolution of the simulation, for reasons which we will 
explore later, the effect is not large, and the simulations converge on very 
similar values for $\Htmass$ after the first couple of  megayears of evolution.
Figure~\ref{H2-MT-fid} therefore demonstrates one of the main results of this 
paper: {\em the timescale for $\mHt$ formation in turbulent gas is much shorter 
than in quiescent gas with the same mean density and large quantities of molecular 
gas can be produced in turbulent regions within only a few 
megayears}. As we will see 
as we proceed, we obtain qualitatively similar results for a wide range of initial 
conditions.

\begin{figure}
\centering
\epsfig{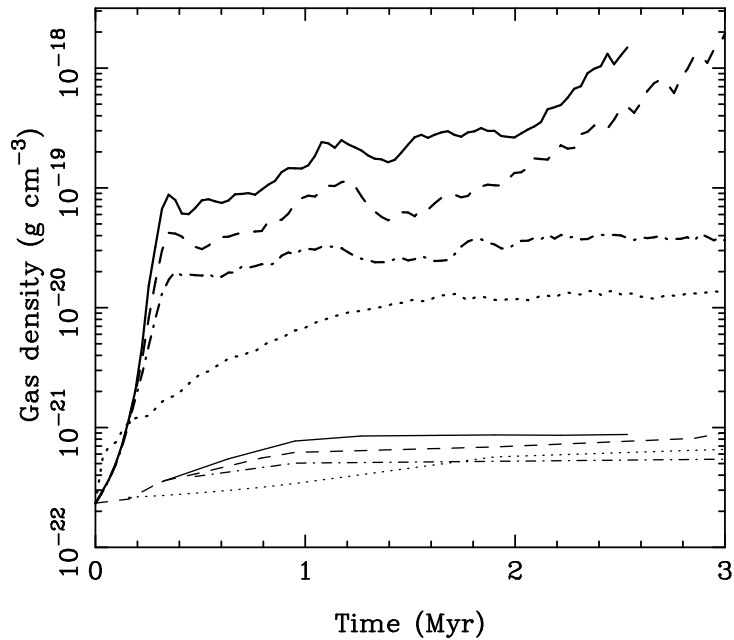}
\caption{Evolution with time of $\rho_{\rm max}$ and $\rho_{\rm rms}$
in four runs with increasing numerical resolution: run MT64 (thick and thin dotted lines),
run MT128 (thick and thin dot-dashed lines), run MT256 (thick and thin dashed lines) 
and run MT512 (thick and thin solid lines). Note that the time sampling of the data used 
to compute $\rho_{\rm rms}$ is ten times coarser than that used to compute 
$\rho_{\rm max}$. \label{dens-MT}}
\end{figure}

Examination of the evolution of the peak density, $\rho_{\rm max}$, 
and the RMS density, $\rho_{\rm rms}$, in our turbulent simulations
(plotted in Figure~\ref{dens-MT} with thick lines and thin lines 
respectively) gives us a strong hint as to why $\Htmass$ grows so rapidly in 
turbulent gas. The turbulence very quickly produces strong compressions in the initially 
uniform density field, significantly increasing the RMS density of the gas within
a megayear, and producing large peak densities. At times $t > 1 \: {\rm Myr}$,
the growth in $\rho_{\rm rms}$ and $\rho_{\rm max}$ slows as the turbulence starts 
to become fully developed, before accelerating again at $t \sim 2 \: {\rm Myr}$ (in the high 
resolution $256^{3}$ and $512^{3}$ runs) or $t \sim 4 \: {\rm Myr}$ (in the lower 
resolution $64^{3}$ and $128^{3}$ runs) due to the onset of runaway gravitational 
collapse.

It is clear from Figure~\ref{dens-MT} that the values of $\rho_{\rm rms}$ and 
$\rho_{\rm max}$ that we obtain from our simulations have not converged, and 
therefore that the density field is not fully resolved even in our $512^{3}$ run.
This is a major cause of the differences apparent in the evolution of $\Htmass$ in
the various runs -- in the higher resolution runs, we resolve more of the dense
structure, and so form more $\mHt$. We would expect this trend to continue as 
we increase the numerical resolution until we reach a point at which the flow is
fully resolved. However, as discussed in \S~4.3.2 of paper I, to fully resolve
shocks at the densities of interest requires a physical resolution of less than
0.001~pc, which we cannot realistically achieve with a fixed grid code. Fortunately,
these resolution issues do not seem to have a large impact on the values we obtain
for $\Htmass$ (other than in run MT64, which is very poorly resolved), suggesting
that our results are not particularly sensitive to the accuracy with which we
can follow the densest parts of the flow. As we shall see in \S~\ref{morph},
this is probably because dense gas in our simulations quickly becomes almost
fully molecular, at which point it clearly ceases to contribute significantly to the
growth of $\Htmass$. Hence, even fairly large errors in the computed densities
in these fully molecular regions have little impact on $\Htmass$.

One obvious question is whether the rapid growth in the value of $\Htmass$ in our 
simulations is driven directly by gravitational collapse, which we expect to 
occur much earlier in a supersonically turbulent medium due to the density enhancements 
created by the turbulence, or whether it is these density enhancements alone that
are responsible for the increased rate of $\mHt$ formation. Figure~\ref{dens-MT} 
suggests that at early times, it is turbulence which plays the major role, with
gravitational collapse becoming important only at later times.

\begin{figure}
\centering
\epsfig{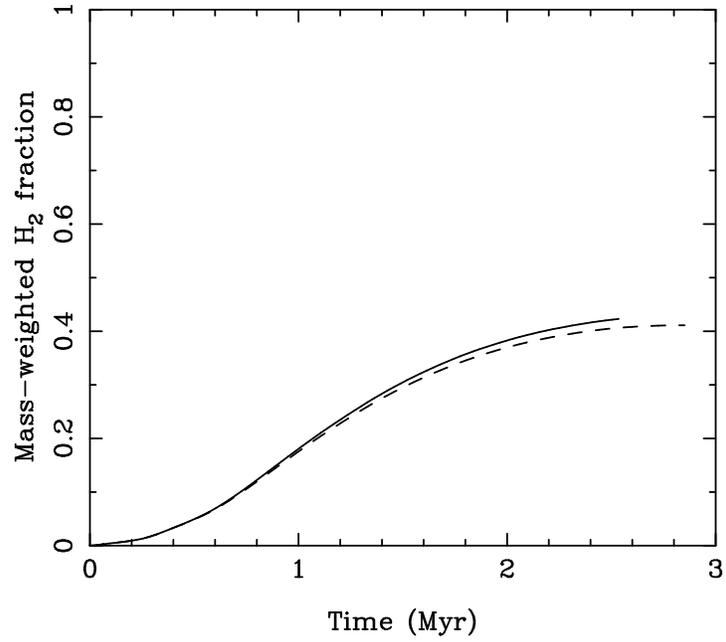}
\caption{Evolution of $\Htmass$ with time in two $512^{3}$ zone runs performed with 
and without self-gravity: run MT512 (solid line) and run MT512-ng (dashed line).
\label{H2-MT-comp}}
\end{figure}

To investigate this point in more detail, we ran a set of simulations in which we disabled 
the effects of self-gravity in the code. The results of these simulations, which we 
denote as MT64-ng, MT128-ng, MT256-ng and MT512-ng are summarized in Table~\ref{xh2_at_end}, 
while in Figure~\ref{H2-MT-comp} we compare the evolution of $\Htmass$ in the highest
resolution simulation, MT512-ng, with its evolution in run MT512. From the figure
it is clear that gravitational collapse is responsible for very little of the $\mHt$
formation seen in the simulations. Most of the $\mHt$ that forms in our turbulent simulations does 
so in dense regions that are {\em not} gravitationally bound. This is an important 
point, as it means that the rapid $\mHt$ formation that we see in our simulations is not
specific to the case of gravitationally bound clouds: we would expect to see a similar 
effect in unbound atomic regions in which significant turbulence is present (such as
the thin, dense sheets formed in the simulations of \citet{ah05} or \citet{vs06}) provided 
that enough gas is present to allow for effective self-shielding of the resulting $\mHt$. 
Indeed, a similar effect has been noted in a different context by \citet{pav02}, who 
simulate the effects of high-speed ($v_{\rm rms} > 15 \: {\rm km} \: {\rm s^{-1}}$),
decaying turbulence in dense gas ($n = 10^{6} \: {\rm cm^{-3}}$), and show that 
following an initial period of dissociation, the $\mHt$ reforms rapidly in the turbulent
gas.

\begin{figure}
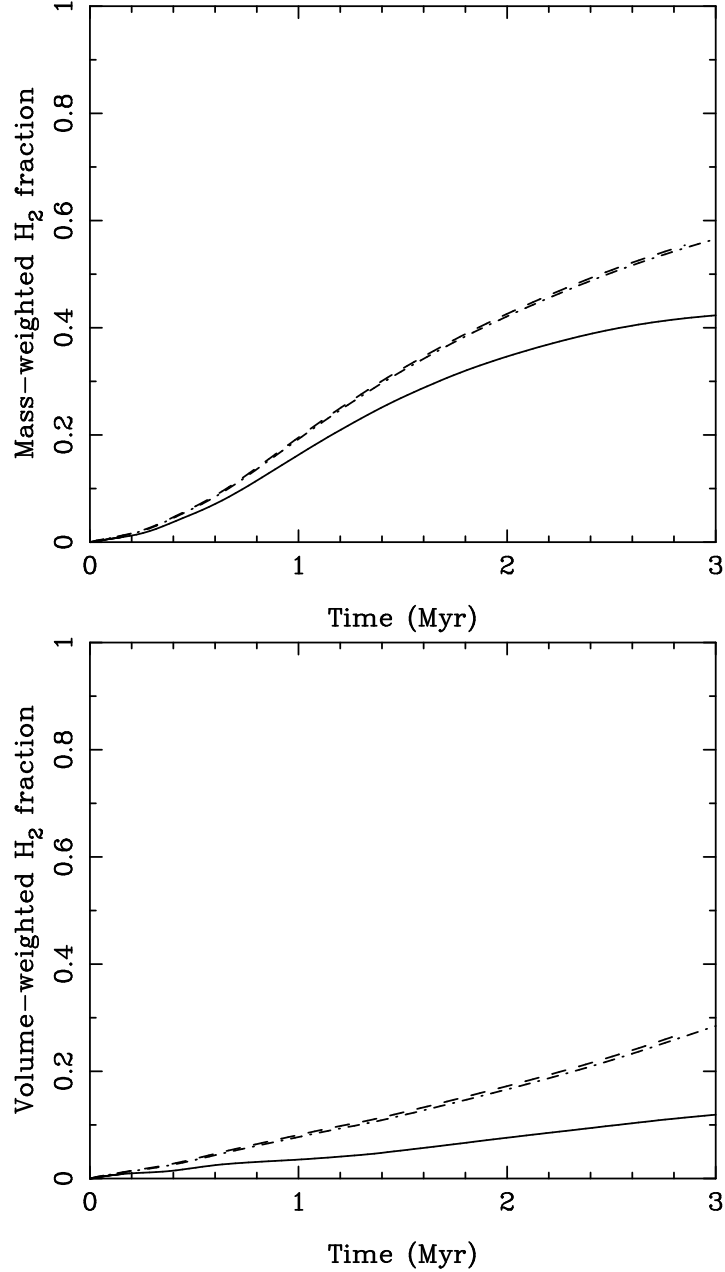

\centering
\epsfig{figure=fig4a.eps,width=20pc,angle=270,clip=}
\epsfig{figure=fig4b.eps,width=20pc,angle=270,clip=}
\caption{(a) Evolution of $\Htmass$ with time in three $256^{3}$ zone 
runs, performed using our fiducial initial conditions. In run MT256
(solid line) $\mHt$ photodissociation was modelled using the local 
shielding approximation, while in run MT256-RT (dashed line) we used
the six-ray shielding approximation. Finally, in run MT256-nr
(dot-dashed line), the strength of the ultraviolet background was set 
to zero.  
Note that the six-ray approximation results follow the zero radiation
results very closely.
(b) As (a), but for the evolution of the volume-weighted molecular
fraction, $\Htvol$. \label{H2-MT-RT}}
\end{figure}

It is also important to establish how sensitive our results are to the approximations
that we have made in order to treat the effects of the UV radiation field. In 
Figure~\ref{H2-MT-RT}a we compare the time evolution of $\Htmass$ in run MT256 
with its evolution in run MT256-RT, a run performed using the same initial conditions
as run MT256, but which used the six-ray shielding approximation instead 
of the local shielding approximation. It is clear that the difference between the 
two runs is small. Slightly more $\mHt$ forms in run MT256-RT, which is just as
we would expect given the greater amount of shielding in that run, but the 
difference between the two runs is no more than 30\% at $t \simeq 3 \: {\rm Myr}$.
In this plot we also show how $\Htmass$ evolves in the absence of an ultraviolet
background, using results from run MT256-nr, which used the same initial conditions
as run MT256, save that $\chi = 0.0$. The results from this run are very similar to
those from run MT256-RT.

This figure demonstrates that for supersonically turbulent flow, our 
approximations work reasonably well. At early times, the rate of growth of 
$\Htmass$ is determined primarily by the time taken to form the $\mHt$, 
which is essentially the same in all three simulations, and so the 
results agree very well. At later times, $\mHt$ photodissociation begins 
to have more of an impact in run MT256, and the results of this run slowly 
diverge from those of runs MT256-RT and MT256-nr. The continued agreement
of the latter runs indicates that there is enough shielding in run MT256-RT
to make $\Htmass$ almost insensitive to the value of $\chi$ (at least for
the range of values examined here). It may be that the results from this
run are closer to the true behaviour than the results of run MT256. 
However, as we cannot be certain of whether the value of $\Htmass$ we 
obtain from run MT256-RT is an overestimate or an underestimate, whereas
we can be certain that the value in run MT256 is an underestimate, we
prefer to focus on the latter results, since they allow us to draw
conclusions that are more robust.

A further point to investigate is how accurately our simulations can follow $\Htvol$, the 
volume-weighted $\mHt$ fraction, defined as
\begin{equation}
\Htvol = \frac{\sum_{i,j,k} x_{\mHt}(i,j,k)}{N},
\end{equation}
where $x_{\mHt}(i,j,k)$ is the $\mHt$ fraction in the grid zone with coordinates $(i,j,k)$,
and where $N$ is the total number of grid zones. In a turbulent simulation, where the 
distribution of mass is highly inhomogeneous (see \S~\ref{morph}), $\Htvol$ will be 
far more sensitive to the behaviour of gas in low-density regions than $\Htmass$.
The evolution of $\Htvol$ in runs MT256, MT256-RT and MT256-nr is plotted in 
Figure~\ref{H2-MT-RT}b. It is clear from the figure that in this case
there is more disagreement between the results obtained with our two different
shielding approximations, and hence more uncertainty in the true value of $\Htvol$.
However, even in this case, the uncertainty is no more than a factor of two at
$t \simeq 3 \: {\rm Myr}$. The fact that we find better agreement for $\Htmass$ than
for $\Htvol$ implies that our approximations do a better job of modelling the behaviour 
of high density, $\mHt$-rich regions than of low density, $\mHt$-poor regions. 
However, this is just what we expect: in regions in which $\xhteq \sim 1$, even large 
errors in $f_{\rm shield}$ or $f_{\rm dust}$ will cause only small errors in the $\mHt$ 
fraction, whereas in regions with $\xhteq \ll 1$, small errors in $f_{\rm shield}$ or 
$f_{\rm dust}$ can lead to large errors in the $\mHt$ fraction. 

Finally, it is necessary to address the question of whether gravitationally unstable regions
in our turbulent simulations are well-resolved. \citet{true97} showed that in order to
properly resolve collapse and avoid artificial fragmentation, it is necessary to resolve 
the local Jeans length of the gas by at least four grid zones. In other words, collapse is 
resolved only while
\begin{equation}
 \Delta x \leq \frac{1}{4} L_{\rm J}(\rho, T),
\end{equation}
where $\Delta x$ is the width of a single grid zone, and $L_{\rm J}$ is
the Jeans length, given by
\begin{equation}
 L_{\rm J} = \frac{\pi^{1/2}c_{\rm s}}{\sqrt{G\rho}},
\end{equation}
where $c_{\rm s}$ is the adiabatic sound speed. Since the densest gas in our simulations
is also the coolest and so has the smallest sound speed, we can determine when (and if)
the Truelove criterion is first violated by following the evolution with time of $L_{\rm J}$ 
in the grid zone with the highest gas density. We have done this for each of our simulations,
and list the results in Table~\ref{xh2_at_end}. We see that in most of our simulations,
the Truelove criterion is violated within the first 1--2~Myr. This would appear to call into 
question our results at later times. However, as we discussed previously in paper I, the fact 
that we no longer properly resolve gravitational collapse in dense gas once the Truelove 
criterion is violated does not necessarily invalidate all of our subsequent results: the key 
question is {\em how much} gas is found in unresolved regions. To quantify this, we
examined intermediate output dumps of density, internal energy and $\mHt$ fraction 
from each of our standard runs, determined which zones were unresolved in each 
case, and computed the fraction of the total gas mass in resolved regions, $f_{\rm res}$, 
and the fraction of the total $\mHt$ mass in the same resolved regions, $f_{\rm res, \mHt}$ 
for every output time for each run. The values of $f_{\rm res}$ and $f_{\rm res, \mHt}$ that
we found at the end of each run are summarized in Table~\ref{fres_at_end}.

We found that in most of our simulations, we resolved zones containing more than 90\%
of the mass and of the $\mHt$ for the whole duration of the simulation, and that in our 
highest resolution run (MT512) we resolved more than 99.8\% of the mass and 99.4\%
of the $\mHt$. The only run in which our resolution was significantly poorer than this 
was our lowest resolution $64^{3}$ run MT64, in which we resolve no more than 50--60\% 
of the gas mass and $\mHt$ mass. We therefore argue that any errors we make in following 
the subsequent evolution of regions that fail to satisfy the Truelove criterion will not 
have a major effect on the evolution of $\Htmass$ and so will not significantly alter our 
main conclusions. 

\section{${\mathbf H_{2}}$ distribution}
\label{mol_distrib}
\subsection{Morphology}
\label{morph}
\begin{figure}
\centering
\epsfig{figure=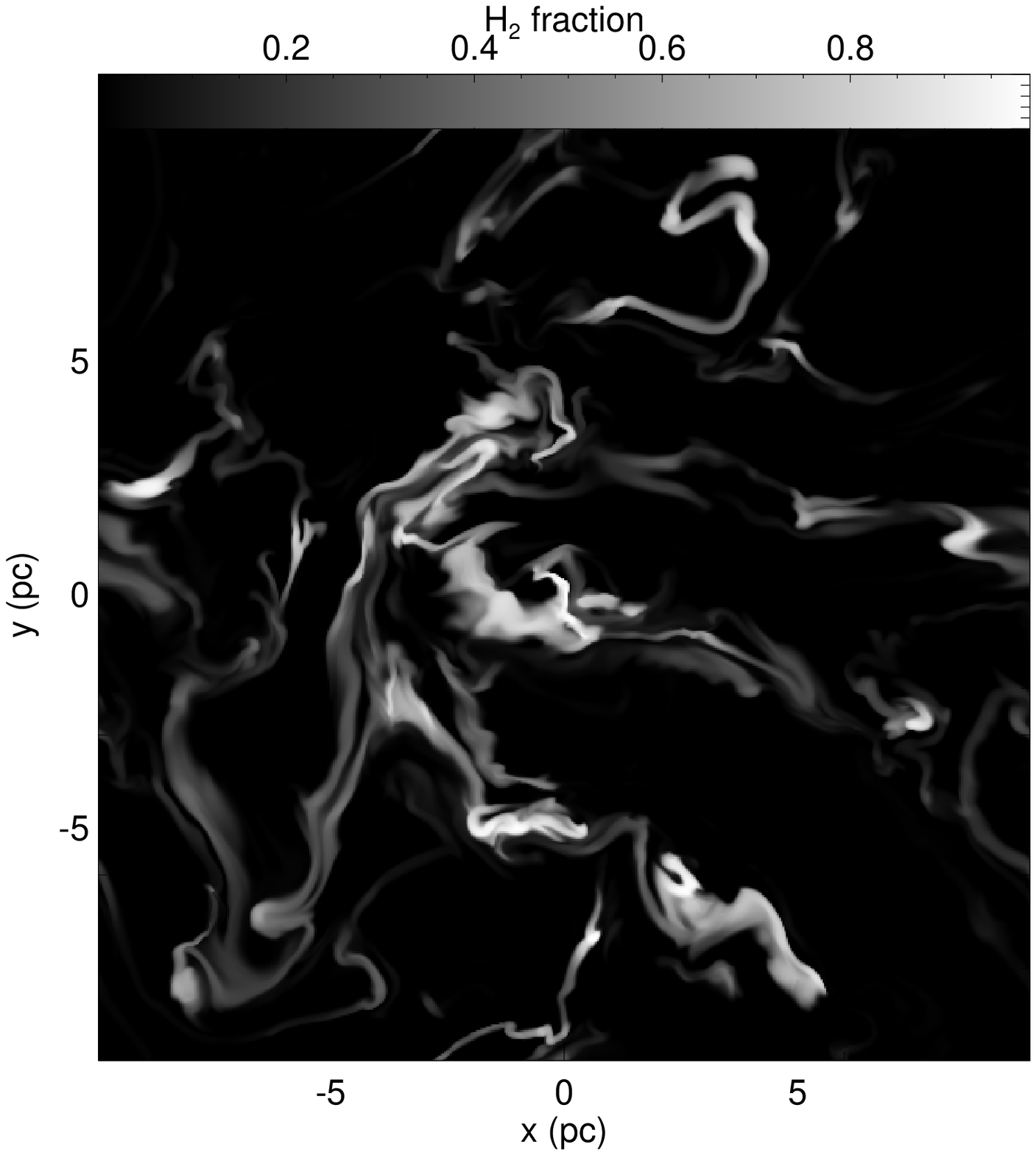,width=20pc,angle=0,clip=}
\epsfig{figure=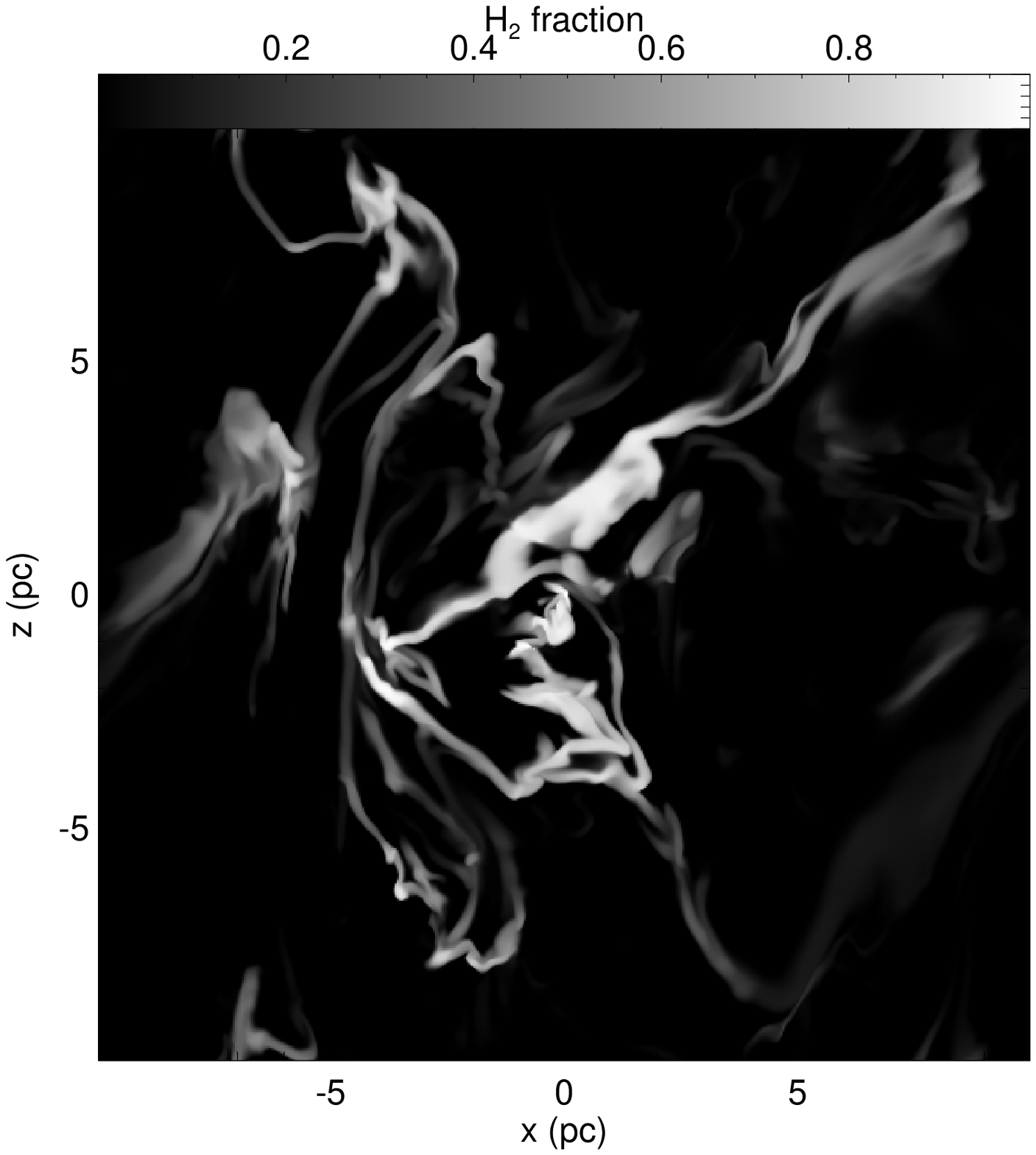,width=20pc,angle=0,clip=}
\caption{(a) Slice in the $x$-$y$ plane through $512^{3}$ zone run MT512 
at time $t = 1.9 \: {\rm Myr}$ showing the spatial variation of the $\mHt$ 
fraction. We have used the fact that our simulations were performed
with periodic boundary conditions to shift the image so that the 
zone with the highest gas density lies at the center of the figure.
(b) As (a), but for a slice in the $x$-$z$ plane. \label{H2-MT-slice}}
\end{figure}

In addition to determining the timescale for $\mHt$ formation, we would also like to know
how the resulting $\mHt$ is distributed in  the gas. We can get an immediate visual
impression of this by looking at how $x_{\mHt}$ varies within a planar slice taken
from the simulation volume. For example, in Figures~\ref{H2-MT-slice}a and 
\ref{H2-MT-slice}b we show slices through run MT512 at time $t = 1.9 \: {\rm Myr}$
in the $x$-$y$ and $x$-$z$ planes respectively. We have made use of the periodic 
boundary conditions to orient the datacube so that both figures are centered on the 
zone with the largest density in the simulation. These figures demonstrate that the 
distribution of molecular gas in the box is highly inhomogeneous, with the largest 
concentrations being found in thin, filamentary structures that fill only a small fraction 
of the total volume. Comparison of the two figures suggests that there is a higher 
degree of structure in the $x$-$z$ plane than in the $x$-$y$, possibly due to the fact 
that the magnetic field is initially aligned parallel to the $z$-axis, but the highly ordered 
structure found in the simulations of magnetized collapse without turbulence analyzed 
in paper I is clearly absent.

\begin{figure}
\centering
\epsfig{figure=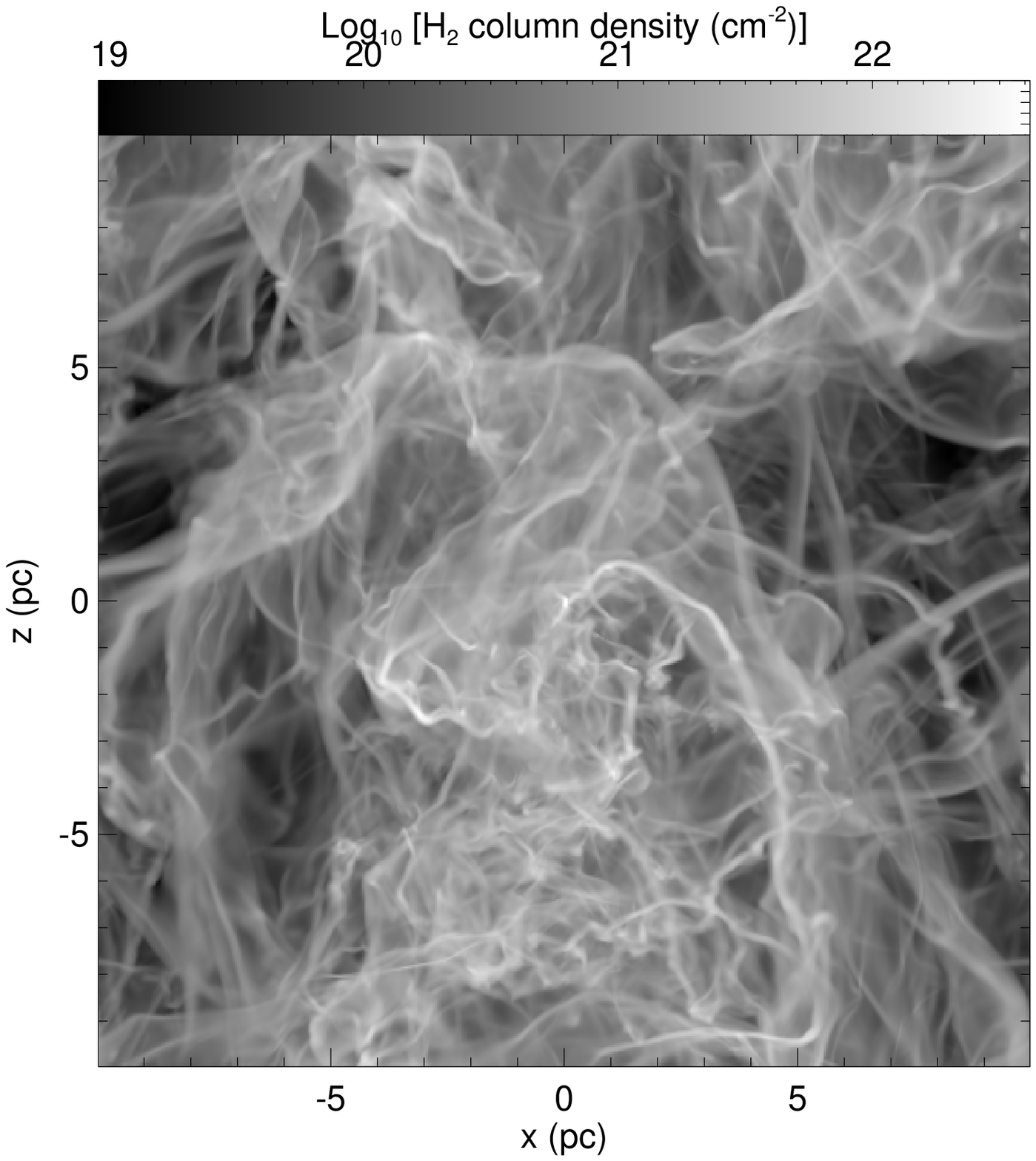,width=20pc,angle=0,clip=}
\epsfig{figure=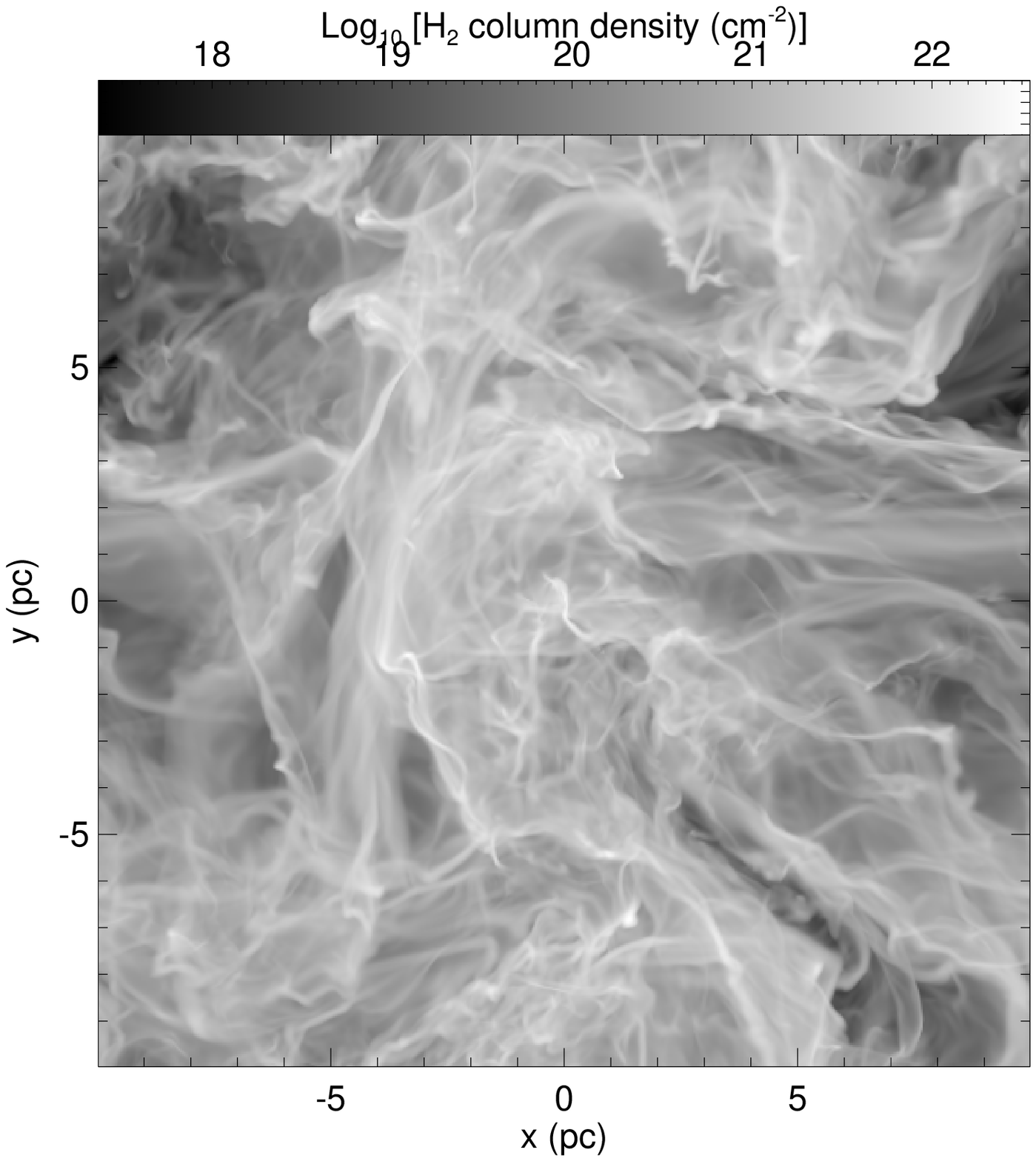,width=20pc,angle=0,clip=}
\figcaption{(a) The $\mHt$ column density in $512^{3}$ zone run MT512 at time 
$t=1.9 \: {\rm Myr}$, computed for lines of sight parallel to the $y$-axis of 
the simulation. (b) As (a), but for lines of sight parallel to the $z$-axis. \label{NH2-MT}}
\end{figure}

An alternative way of visualizing the $\mHt$ distribution is to look at the molecular 
gas in projection, by computing the $\mHt$ column density distribution that would 
be seen by a distant observer. In Figures~\ref{NH2-MT}a and \ref{NH2-MT}b, we show
the $\mHt$ column density distribution that would be seen by an observer looking along
the $y$ axis or the $z$ axis respectively. These images again highlight the inhomogeneous
nature of the $\mHt$ distribution, and also demonstrate that considerable small-scale
structure exists in this distribution, despite the fact that our initial turbulent 
velocity field included power only on large scales. 

\subsection{Density dependence of the ${\mathbf H_{2}}$ fraction}
\label{xh2n}
\begin{figure}
\centering
\epsfig{figure=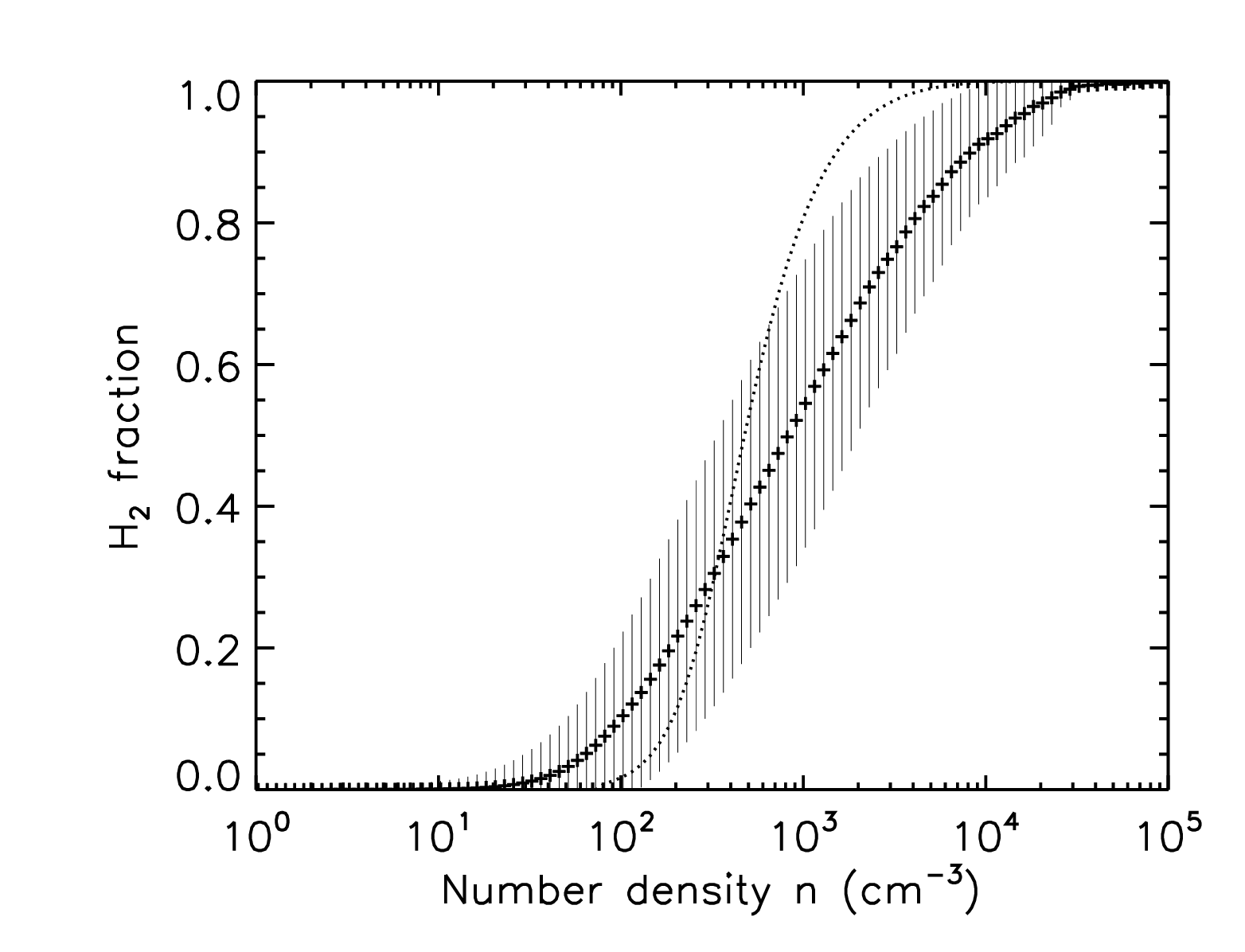,width=25pc,angle=0,clip=}
\caption{$\mHt$ fraction as a function of the number density of the
gas (crosses) for $512^{3}$ zone run MT512 at time $t = 1.9 \: {\rm Myr}$. To compute 
these values, we binned the data by number density, using bins of width 0.05 dex, 
and computed the mean value of $x_{\mHt}$ for each bin. The standard deviation 
in the value of $x_{\mHt}$ in each bin is also indicated, as is the mean value 
of $\xhteq$ for each bin (dotted line).}
\label{xh2vn-MT}
\end{figure}

We can study the $\mHt$ distribution in a more quantitative fashion by examining how
the $\mHt$ fraction varies with density. To do this, we computed $x_{\mHt}$ and $n$ 
for each of the zones in run MT512 at $t = 1.9 \: {\rm Myr}$ and then binned the data 
by number density, using bins of width 0.05 dex. We then computed the mean and 
standard deviation for $x_{\mHt}$ in each bin. The resulting values are plotted in
Figure~\ref{xh2vn-MT}. Note that although the mean values we compute here are 
volume weighted, we would not expect the mass weighted values to differ greatly, 
since the narrow width of our density bins means that there is little variation in the gas 
mass from zone to zone within a given bin. We also plot in Figure~\ref{xh2vn-MT} the
mean value of  $\xhteq$ for each density bin, computed using the local
shielding approximation.

Three features of Figure~\ref{xh2vn-MT} deserve particular comment. First, it is
clear that there is considerable scatter in the value of $x_{\mHt}$ at a given density.
Comparison with the plots of $x_{\mHt}$ versus $n$ presented in paper I for 
smoothly collapsing gas shows that the scatter is much greater in the turbulent case.
Despite this, however, there is still an obvious underlying trend in the distribution 
of $x_{\mHt}$ with $n$, which runs in the direction that we would expect (i.e.\ high 
density gas is more highly molecular than low density gas). Second, even though 
$\Htmass \simeq 0.4$ at this point in the simulation, there are already regions 
where the $\mHt$ fraction is very much higher, and indeed gas with a number 
density $n > 10^{4} \: {\rm cm^{-3}}$ is already almost entirely molecular. 

\begin{figure}
\centering
\epsfig{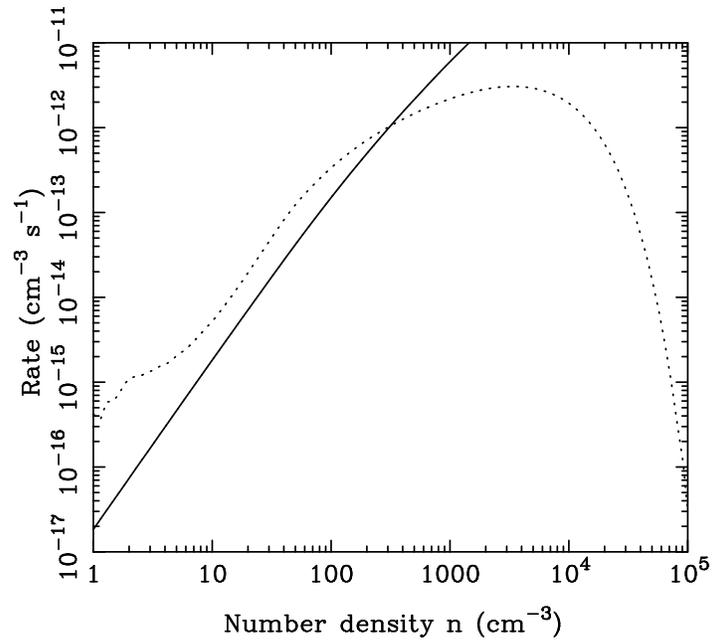}
\caption{Mean $\mHt$ formation rate (solid line) and $\mHt$ photodissociation
rate (dotted line), plotted as a function of density, for $512^{3}$ zone run MT512 at time 
$t = 1.9 \: {\rm Myr}$. For $n < 300 \: {\rm cm^{-3}}$, the photodissociation rate exceeds
the formation rate, indicating that more $\mHt$ is being destroyed in this gas than is
being formed {\em in situ}.}
\label{dxh2-MT}
\end{figure}

Finally, comparison of the distribution of $x_{\mHt}$ with $n$ with the equilibrium 
distribution (computed using the same local shielding approximation as is used in
the simulation, and given by the dashed line in Figure~\ref{xh2vn-MT}) indicates that in most 
of the gas, the $\mHt$ fraction has yet to reach equilibrium. For densities in the range
$300 < n < 10^{4} \: {\rm cm^{-3}}$, the actual $\mHt$ fraction lies below the
equilibrium curve, indicating that the $\mHt$ fraction in this gas is still growing. More
interestingly, in gas with $n < 300 \: {\rm cm^{-3}}$, the mean $x_{\mHt}$ lies {\em above}
the equilibrium curve, implying that the $\mHt$ fraction in this gas must be decreasing:
in other words, more $\mHt$ is being destroyed by photodissociation than is being formed on 
dust grains. To confirm that this interpretation is correct, we have computed the values of
the formation and photodissociation rates of $\mHt$ for each grid zone in run MT512 at
$t = 1.9 \: {\rm Myr}$. The resulting values were binned in a similar fashion to the $\mHt$ 
abundance, and the mean values in each density bin are plotted in Figure~\ref{dxh2-MT}. 
As we can see, the formation rate is larger than the photodissociation rate in gas with 
$n > 300 \: {\rm cm^{-3}}$, but is smaller than the photodissociation rate in gas with
$n < 300 \: {\rm cm^{-3}}$.

\begin{figure}
\centering
\epsfig{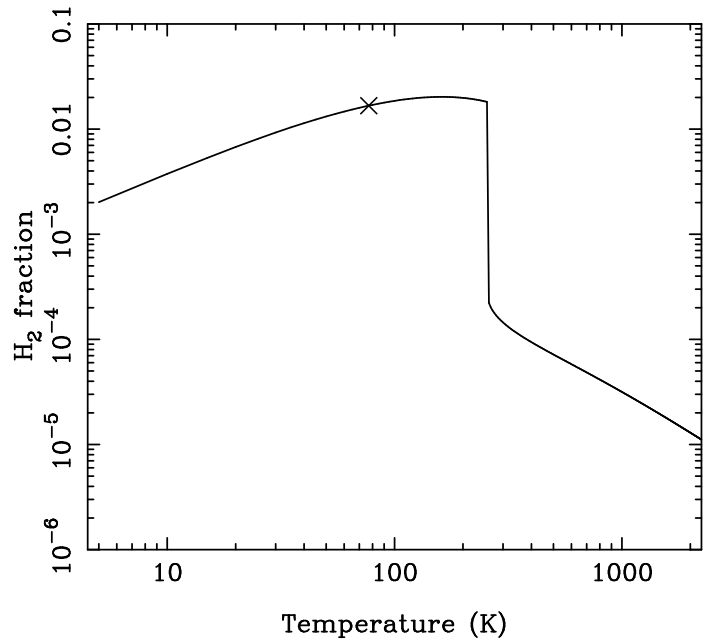}
\caption{Value of $\xhteq$ as a function of the gas temperature $T$, computed for gas
with $n = 100 \: {\rm cm^{-3}}$ which is shielded by a total hydrogen column density
$N_{\rm \mH, tot} = 6.0 \times 10^{18} \: {\rm cm^{-2}}$. Note that this is the same amount 
of shielding as there would be in a zone with this value of $n$ in one of our $512^{3}$, 
$L = 20 \: {\rm pc}$ simulations. The sharp transition in $\xhteq$ seen at 
$T \sim 250 \: {\rm K}$ is due to the sudden onset of self-shielding: at higher temperatures,
there is not enough $\mHt$ in the gas, even in the equilibrium state, to make the Lyman-Werner
bands optically thick.}
\label{xh2eq-vT}
\end{figure}

Since our simulations start with fully atomic gas, there are only two ways to explain the
fact that $x_{\mHt}$ exceeds $\xhteq$ at low densities: either $\xhteq$ must have been
much higher in the past, or some (or all) of the molecular content of the low density gas
must have actually formed in higher density gas and then been transported to lower 
densities by the action of the turbulence. Of these two explanations, the first does not 
appear to be viable -- most of the parameters that determine the value of $\xhteq$ 
at a given density in our simulations (such as the strength of the UV background) 
are fixed, and the one parameter that does vary, the gas temperature, cannot give
us the effect we seek. To demonstrate why, we have plotted in Figure~\ref{xh2eq-vT} the 
variation of $\xhteq$ with temperature for gas with a  number density $n = 100 \: {\rm cm^{-3}}$.
The mean temperature of the $100 \: {\rm cm}^{-3}$ gas in our simulation is approximately
$77 \: {\rm K}$, and the corresponding value of $\xhteq$ is indicated in Figure~\ref{xh2eq-vT} 
by a cross. From the figure it is clear that if the temperature were slightly higher at
early times in the simulation (i.e.\ if $T \sim 200$--300~K), then $\xhteq$ may also have been
slightly higher, but at most by about 20\%. On the hand, if the temperature had been much 
higher, or much lower, then $\xhteq$ would have been smaller than its present value. Given that
we find a mean value of $x_{\mHt}$ at $n = 100 \: {\rm cm^{-3}}$ that is actually five times
greater than the equilibrium value, it is clear that temperature variations at earlier times
can be responsible for no more than a small portion of this excess, and that most or all of
it must therefore be due to the transport of $\mHt$ from high density regions to low density
regions.

\begin{figure}
\centering
\epsfig{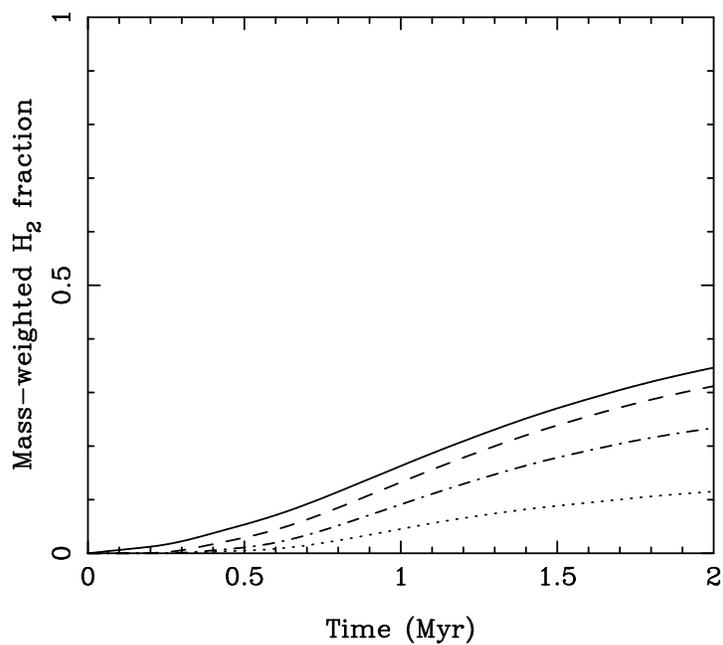}
\caption{Evolution of $\Htmass$ with time in $256^{3}$ zone runs MT256 
(solid line), MT256-th3e2 (dashed line), MT256-th1e3 (dot-dashed line) and 
MT256-th3e3 (dotted line). In runs MT256-th3e2, MT256-th1e3 and MT256-th3e3,
$\mHt$ formation was artificially suppressed in gas with a number density 
$n < n_{\rm th} = 300$, 1000, and $3000 \: {\rm cm^{-3}}$, respectively.}
\label{H2-MT-nth}
\end{figure}

To test this hypothesis, we performed several simulations at $256^{3}$ resolution
in which we artifically suppressed $\mHt$ formation in gas with a number density smaller
than some threshold value $n_{\rm th}$. In runs MT256-th3e2, MT256-th1e3 and
MT256-th3e3 we used values for $n_{\rm th}$ of 300, 1000, and $3000 \: {\rm cm^{-3}}$ 
respectively. Our standard set of input parameters were used for all three runs. In 
Figure~\ref{H2-MT-nth}, we show how $\Htmass$ evolves in these runs; we also plot the 
corresponding evolutionary track from run MT256 for comparison. Unsurprisingly, we see 
that as we increase $n_{\rm th}$, the total amount of $\mHt$ that forms in the 
simulations decreases. However, the reduction in $\Htmass$ is less than a factor of two
for $n_{\rm th} \leq 1000 \: {\rm cm^{-3}}$, suggesting that most of the $\mHt$
that forms (and persists) in the simulation does so in the high density gas.

\begin{figure}
\centering
\epsfig{figure=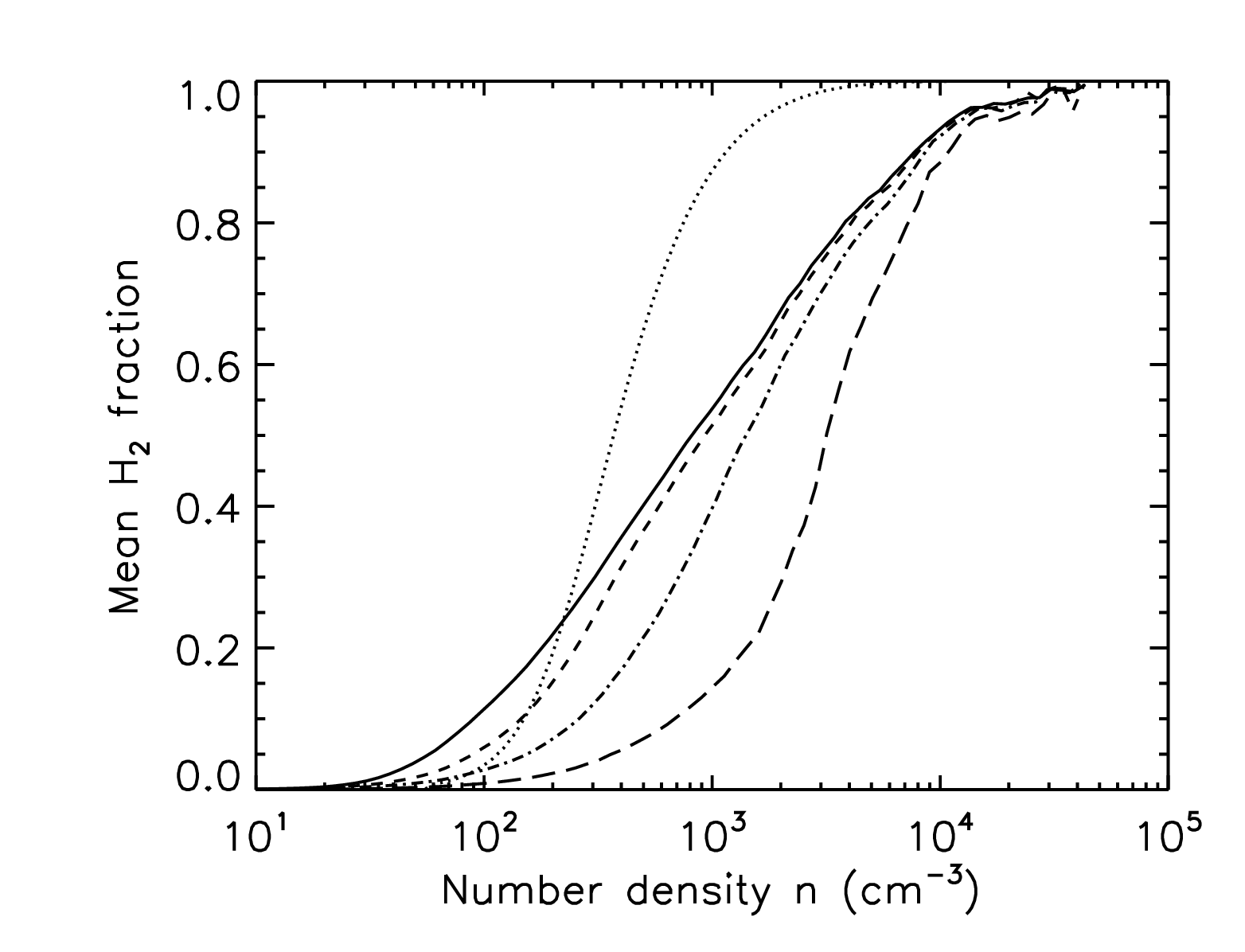,width=25pc,angle=0,clip=}
\caption{$\mHt$ fraction as a function of gas number density
at $t = 1.9 \: {\rm Myr}$ in $256^{3}$ zone runs MT256 (solid line), 
MT256-th3e2 (short-dashed line), MT256-th1e3 (dot-dashed line) and 
MT256-th3e3 (long-dashed line). No $\mHt$ can form in these runs
at densities $n < n_{\rm th}$, where $n_{\rm th} = 0$, 300, 1000 and
$3000 \: {\rm cm^{-3}}$, respectively, which means that $\mHt$ present
at $n < n_{\rm th}$ must have initially formed in denser gas. For reference, 
we also indicate the mean value of $\xhteq$ as a function of number density 
in run MT256 (dotted line).}
\label{xh2vn-MT-nth}
\end{figure}

To examine how the $\mHt$ in these simulations is distributed with density, we computed 
the mean value of $x_{\mHt}$ in a set of density bins of width 0.05 dex in all three 
runs at time $t = 1.9 \: {\rm Myr}$, and compared the values with those previously
computed for run MT256. The results 
are plotted in Figure~\ref{xh2vn-MT-nth}; for clarity, we do not indicate the size of 
the standard deviations in these values, although they are comparable to those in 
Figure~\ref{xh2vn-MT}.

There are two important conclusions that we can draw from this figure. First, it is
clear that although the imposition of a density threshold for $\mHt$ formation 
substantially reduces the amount of $\mHt$ present in gas with $n < n_{\rm th}$, 
it does not eliminate it: some $\mHt$ remains, and since it cannot have formed 
{\em in situ} in these simulations, it must have formed in higher density gas and 
been transported to the low density gas by the turbulent  velocity field. Second, 
imposing a density threshold also affects the $\mHt$ abundance at densities 
above the threshold: although the results of the various runs eventually converge 
at high density, there are clear differences for densities $n \simless 3 n_{\rm th}$, 
rather than at $n \simless n_{\rm th}$ as we might have initially expected. 

In retrospect, this behaviour should perhaps have been expected. Since the
$\mHt$ formation rate scales as the square of the number density, $n^{2}$,
overdense gas makes a disproportionately large contribution to the overall
$\mHt$ formation rate compared to underdense gas or gas with densities
near the mean value. However, previous numerical work on isothermal turbulence 
has shown that many of the overdense structures formed in supersonically 
turbulent flows are transient objects with short lifetimes 
\citep[see e.g.][]{khm00,vsks05}.
Therefore, any $\mHt$ formed within these transient overdensities will not
remain at high densities for long, but will quite quickly find itself carried into
much lower density surroundings.  The results of runs MT256-th3e2, 
MT256-th1e3 and MT256-th3e3 suggest that a large fraction of the $\mHt$ 
in the gas is produced in this fashion.

It would clearly be of great interest to determine the extent to which $\mHt$ 
formed in transient overdensities is actually mixed into lower density material 
as the overdensities expand and are broken up by the flow; i.e.\ the amount
of turbulent mixing of material which occurs. Unfortunately, to determine this 
with an Eulerian code such as ZEUS-MP, we would need to use tracer 
particles to follow individual fluid elements and these are not currently 
implemented in our version of the code. We intend to revisit this issue in 
future work.

\begin{figure}
\centering
\epsfig{figure=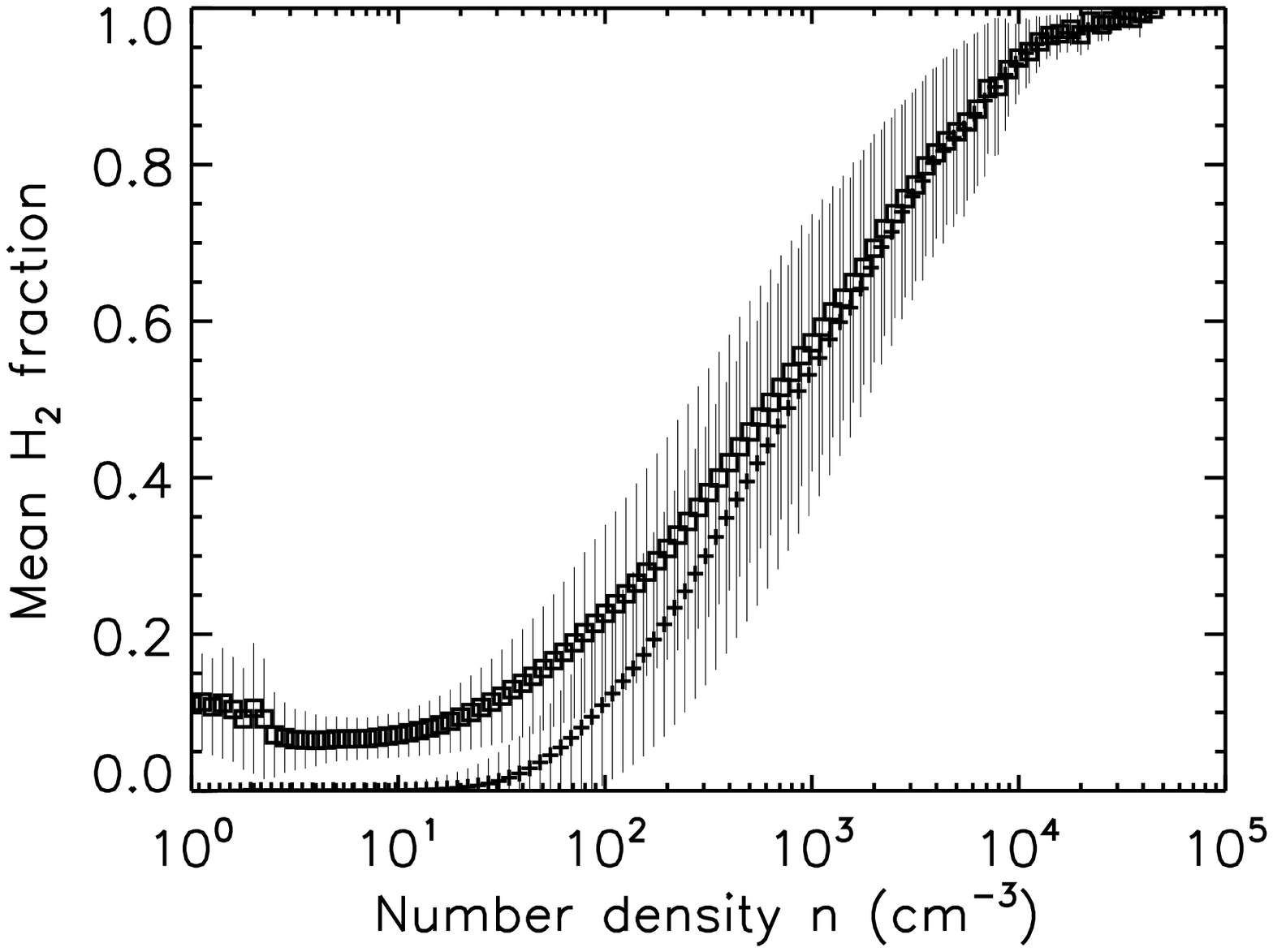,width=25pc,angle=0,clip=}
\caption{
$\mHt$ fraction as a function of the number density of the
gas for $256^{3}$ zone runs MT256 (crosses) and MT256-RT (squares) 
at time $t = 1.9 \: {\rm Myr}$. The standard deviation 
in the value of $x_{\mHt}$ in each bin is also indicated.
\label{xh2vn-MT-RT}}
\end{figure}

Finally, we have investigated how sensitive our results on the density 
distribution of $\mHt$ are to the choice of shielding approximation. In
Figure~\ref{xh2vn-MT-RT} we plot $x_{\mHt}$ as a function of $n$ in runs 
MT256 and MT256-RT at time $t = 1.9 \: {\rm Myr}$. Above $n = 1000 \: {\rm cm^{-3}}$,
the results of these two runs are almost indistinguishable. Since,
as we have just seen, the majority of $\mHt$ in these simulations forms
at densities $n > 1000 \: {\rm cm^{-3}}$, this provides a simple 
explanation for the insensitivity of $\Htmass$ to our choice of shielding
approximation. At lower densities, the difference between runs MT256 and
MT256-RT is greater. However, for $100 < n < 1000 \: {\rm cm^{-3}}$,
the mean value of $x_{\mHt}$ derived in run MT256-RT lies within the range 
of values found in run MT256, and the two runs only differ significantly
at $n < 100 \: {\rm cm^{-3}}$. In run MT256, the $\mHt$ fraction at 
$n < 100 \: {\rm cm^{-3}}$ very quickly falls to almost zero, as there is
not enough shielding at these densities to maintain a large $\mHt$ 
fraction. In run MT256-RT, on the other hand, shielding by the 
surrounding gas enables $\mHt$ to survive
in some regions, even  
at these low densities, producing the extended tail seen in the distribution.

\subsection{Density probability distribution function and cumulative mass function} 
\begin{figure}
\centering
\epsfig{figure=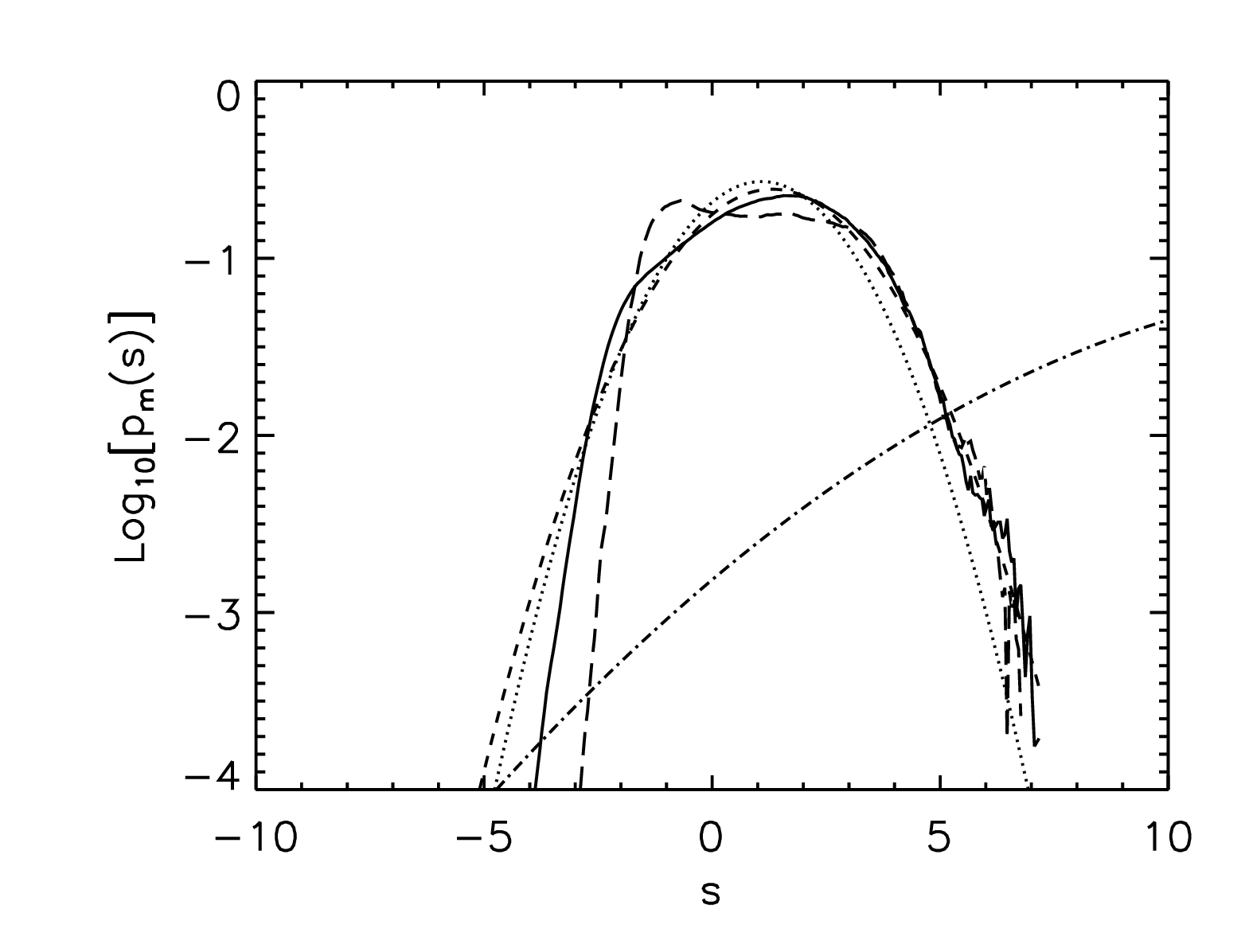,width=25pc,angle=0,clip=}
\caption{Mass-weighted density PDF of the gas in $512^{3}$ zone run MT512 at times
$t = 1.3 \: {\rm Myr}$ (long-dashed line) and $t = 1.9 \: {\rm Myr}$ (solid line). 
Several log-normal functions are also plotted: one with the same mean as the 
$t = 1.9 \: {\rm Myr}$ data, $s_{\rm m} = 1.32$ (short-dashed line), a second with a mean 
$s_{\rm m} = 1.082$, corresponding to the \citet{pnj97} prediction (dotted line) and 
a third with a mean $s_{\rm m} = 15.40$, corresponding to the \citet{pvs98} prediction 
(dot-dashed line). \label{den-pdf-MT} }
\end{figure}

We can further characterize the $\mHt$ distribution produced in our simulations by
determining at which densities the bulk of the molecular gas resides. We have already
seen that the highest molecular fractions are found in the densest gas, but in order
to quantify how much of the total molecular mass is located at high densities, we
need to characterize the density distribution of the gas. To do this, we have computed
the mass-weighted density probability function (PDF) for run MT512 at $t = 1.9 \: 
{\rm Myr}$ (plotted in Figure~\ref{den-pdf-MT}). For ease of comparison with other
values in the literature, we plot the PDF here as a function of $s =  \ln(n / n_{\rm i})$,
where $n_{\rm i} = 100 \: {\rm cm^{-3}}$ is the mean number density of hydrogen nuclei,
which in our simulations is identically equal to the initial number density of hydrogen 
nuclei. We also plot in Figure~\ref{den-pdf-MT} a log-normal function of the form
\begin{equation}
 p_{\rm m}(s) = \frac{1}{\sqrt{2\pi \sigma^{2}}} \exp 
 \left[-\frac{(s - s_{\rm m})^{2}}{2 \sigma^{2}} \right], \label{logn-pdf}
\end{equation}
where $s_{\rm m}$ and $\sigma$ are the mean and dispersion of the log-normal
distribution, respectively; note that these are related by $|s_{\rm m}| = \sigma^{2}/2$. 
To construct the log-normal distribution in Figure~\ref{den-pdf-MT} we have set 
$s_{\rm m} = 1.32$, which is the mean of the actual dataset. Figure~\ref{den-pdf-MT}
demonstrates that at densities above the mean ($s > 0$), the density PDF of run
MT512 is close to a log-normal in form. On the other hand, at densities below the
mean, clear deviations from a log-normal are apparent: there is a pronounced 
excess of power at $s = -2$, corresponding to $n \simeq 10 \: {\rm cm^{-3}}$, and
a deficit of power at lower densities. Previous investigations by \citet{pvs98}
and \citet{svcp98} of the influence of the polytropic exponent $\gamma$ on the shape 
of the density PDF produced by a turbulent flow suggested that a highly turbulent flow 
with $\gamma < 1$ should produce a power-law density PDF at densities above the mean, and 
that a log-normal PDF is only recovered if $\gamma = 1$; i.e.\ if the gas is isothermal. 
However, this prediction is based primarily on the results from one-dimensional 
simulations and the predicted behaviour is not seen in the three-dimensional simulations 
of \citet{lkm03} or \citet{mbka05}. Our simulations have an effective polytropic exponent 
$\gamma_{\rm eff} < 1$ (see \S~\ref{gamma} below), but we also see no convincing 
evidence for power-law behaviour at high density, suggesting that in this case the results of 
one-dimensional simulations are also not a good guide to the three-dimensional behaviour.

For driven {\em isothermal} turbulence, there is general agreement that the resulting density
PDF should have a log-normal form \citep{pnj97,pvs98,np99}. However, there is less agreement
on the dispersion of this PDF. Based on the results of three-dimensional turbulence simulations,
\citet{pnj97} predict that the dispersion, $\sigma$, is related to the RMS Mach number of the 
flow, ${\cal M}$, by
\begin{equation}
 \sigma^{2} = \ln \left(1 + 0.25 {\cal M}^{2} \right).
\end{equation}
On the other hand, based on their one-dimensional simulations,
\citet{pvs98} predict that the relationship between $\sigma$ and
${\cal M}$ is actually $ \sigma \simeq {\cal M}$.  At $t = 1.9 \: {\rm
Myr}$, the RMS Mach number in run MT512 is ${\cal M} = 5.55$, and so
according to the \citeauthor{pnj97} prediction, we should find that
$\sigma = 1.471$, and hence that $s_{\rm m} = 1.082$. On the other
hand, the \citeauthor{pvs98} prediction gives $\sigma = 5.55$ and
hence $s_{\rm m} = 15.40$. In Figure~\ref{den-pdf-MT}, we compare
these predictions with the actual density PDF in run MT512 at this
time. We see that although neither prediction fits the data perfectly,
the \citeauthor{pnj97} prediction comes very much closer to doing so
than the \citet{pvs98} prediction; the former merely underestimates
the amount of dense gas slightly, while the latter dramatically
overestimates it.

Since \citeauthor{pnj97} simulate driven turbulence in isothermal gas without self-gravity,
whereas we simulate decaying turbulence in non-isothermal gas with self-gravity, it is
perhaps not surprising that we obtain a slightly different result. We have examined whether
the difference that we find between the \citeauthor{pnj97} prediction and the true PDF is
due to the inclusion of self-gravity in our simulations by comparing the density PDF we
obtained from run MT512 at $t=1.9 \: {\rm Myr}$ with that in run
MT512-ng at the same time. We find no significant difference,
suggesting that self-gravity has so far had little 
effect on the PDF. The difference may rather be due to the fact that we are examining 
decaying rather than driven turbulence, since the highest density portion of the PDF appears 
to be established at an earlier time than the one which we examine here, as can be seen
from comparison of the density PDF at $t = 1.3 \: {\rm Myr}$ (the long-dashed line in 
Figure~\ref{den-pdf-MT}) with that at $t = 1.9 \: {\rm Myr}$. Alternatively, the difference 
may be due to our softer equation of state.  
\citet{mbka05} find a slightly higher level of disagreement between
the dispersions of the density PDFs in their 
simulations and the predictions of \citeauthor{pnj97}, using highly 
non-isothermal turbulent gas to model the larger-scale ISM. Further
investigation of this point might be worthwhile.  
Nevertheless, whatever the reason for the residual differences, it is clear that the 
\citeauthor{pnj97} prescription provides a good description of the density 
PDF of the gas, while the \citeauthor{pvs98} prescription does not.

As far as the distribution of molecular gas is concerned, the main lesson to
be drawn from Figure~\ref{den-pdf-MT} is that only a small fraction of the gas is 
found at densities $n > 10^{4} \: {\rm cm^{-3}}$. Therefore, even though this material is fully 
molecular, it does not represent a large fraction of the total mass of $\mHt$. Instead, most 
of the $\mHt$ is found in regions with number densities in the range 
$10^{2} \leq n \leq 10^{4} \: {\rm cm^{-3}}$, as can be seen more easily by examining the 
behaviour of the cumulative mass distribution of $\mHt$,  plotted in 
Figure~\ref{xh2vn-MT-times}b below.

\begin{figure}
\centering
\epsfig{figure=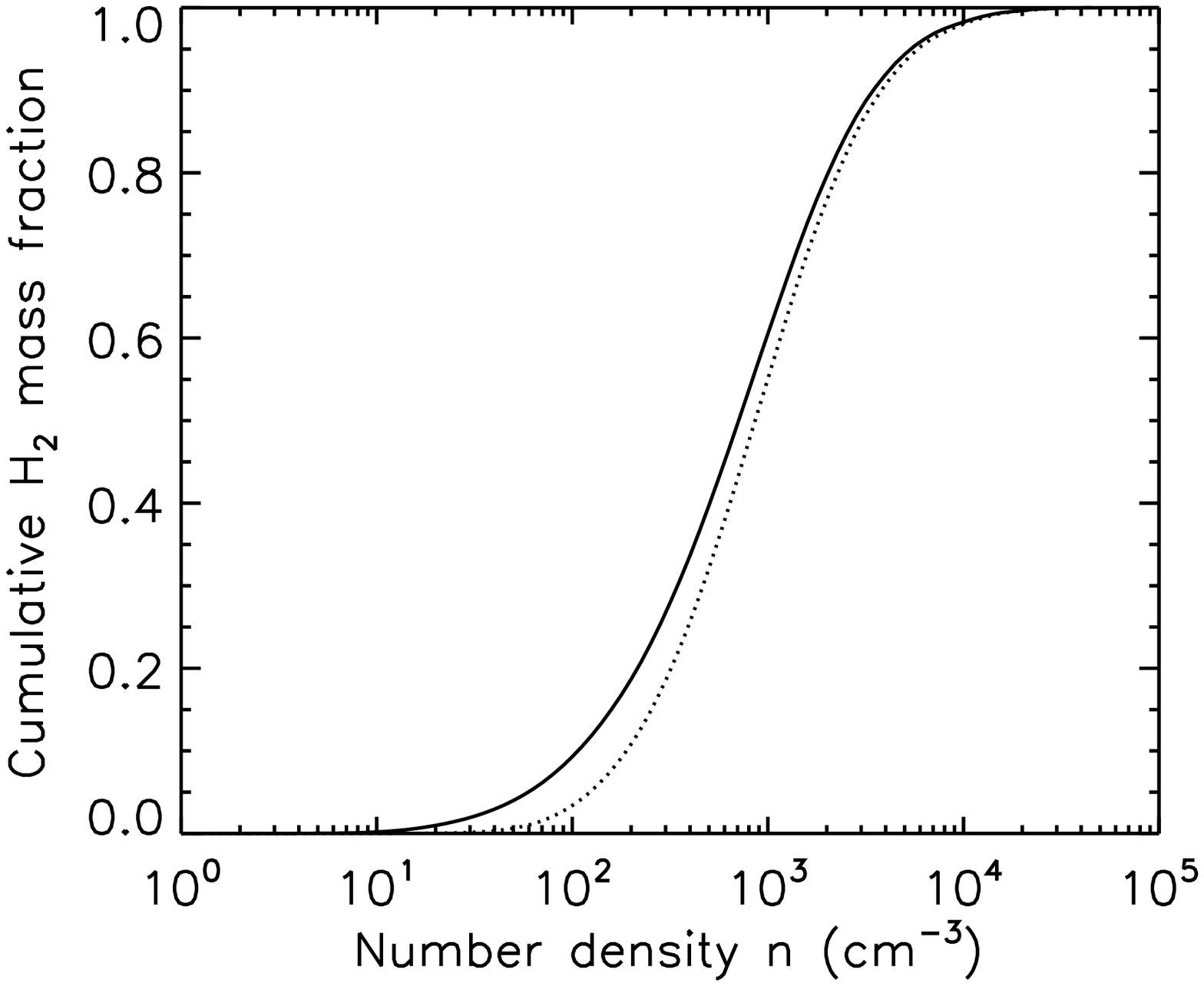,width=25pc,angle=0,clip=}
\caption{Cumulative mass distribution of $\mHt$ with $n$ in 
$256^{3}$ zone runs 
using the six-zone approximation (MT256-RT; {\em solid line}) and the local
approximation (MT256; {\em dotted line})
at time $t = 1.9 \: {\rm Myr}$. \label{cuml-H2-RT}}
\end{figure}

Finally, as part of our effort to understand the sensitivity of our results to our
treatment of UV shielding, we have compared the cumulative mass distribution 
of $\mHt$ in runs MT256 and MT256-RT, performed using the local shielding 
approximation and the six-ray shielding approximation respectively. The results
are shown in Figure~\ref{cuml-H2-RT}. We see that although slightly more of the total 
mass of $\mHt$ is found at low densities in run MT256-RT than in run MT256, the 
difference is small, suggesting that the results discussed above are not significantly 
affected by our choice of shielding approximation.

\subsection{Morphological evolution}
So far, we have focused primarily on understanding the distribution of $\mHt$ in our 
simulations at a single moment in time. The particular time that we have chosen, 
$t = 1.9 \: {\rm Myr}$,
corresponds to slightly less than a turbulent crossing time at our initial RMS turbulent 
velocity, and so represents a point in time at which the turbulence in the simulations is already 
well-developed, but has not yet decayed to insignificance.  Moreover, it is also early enough 
that we can be reasonably confident that even in those cases where the runaway gravitational 
collapse of one or more dense clumps is occurring in the cloud, widespread star formation 
has not yet had sufficient time to occur,  and therefore that our neglect of star formation and 
the associated feedback processes does not lead us to derive misleading results. 

However, the choice of this particular moment in time remains somewhat arbitrary, and it is
also not immediately obvious how the $\mHt$ morphology of the gas at this point in time is
related to the $\mHt$ morphology at earlier or later times. To investigate this, we have 
therefore examined the distribution of $\mHt$ in run MT512 at several different output times 
between $t = 0.6 \: {\rm Myr}$ and $t = 2.5 \: {\rm Myr}$.

\begin{figure}
\centering
\epsfig{figure=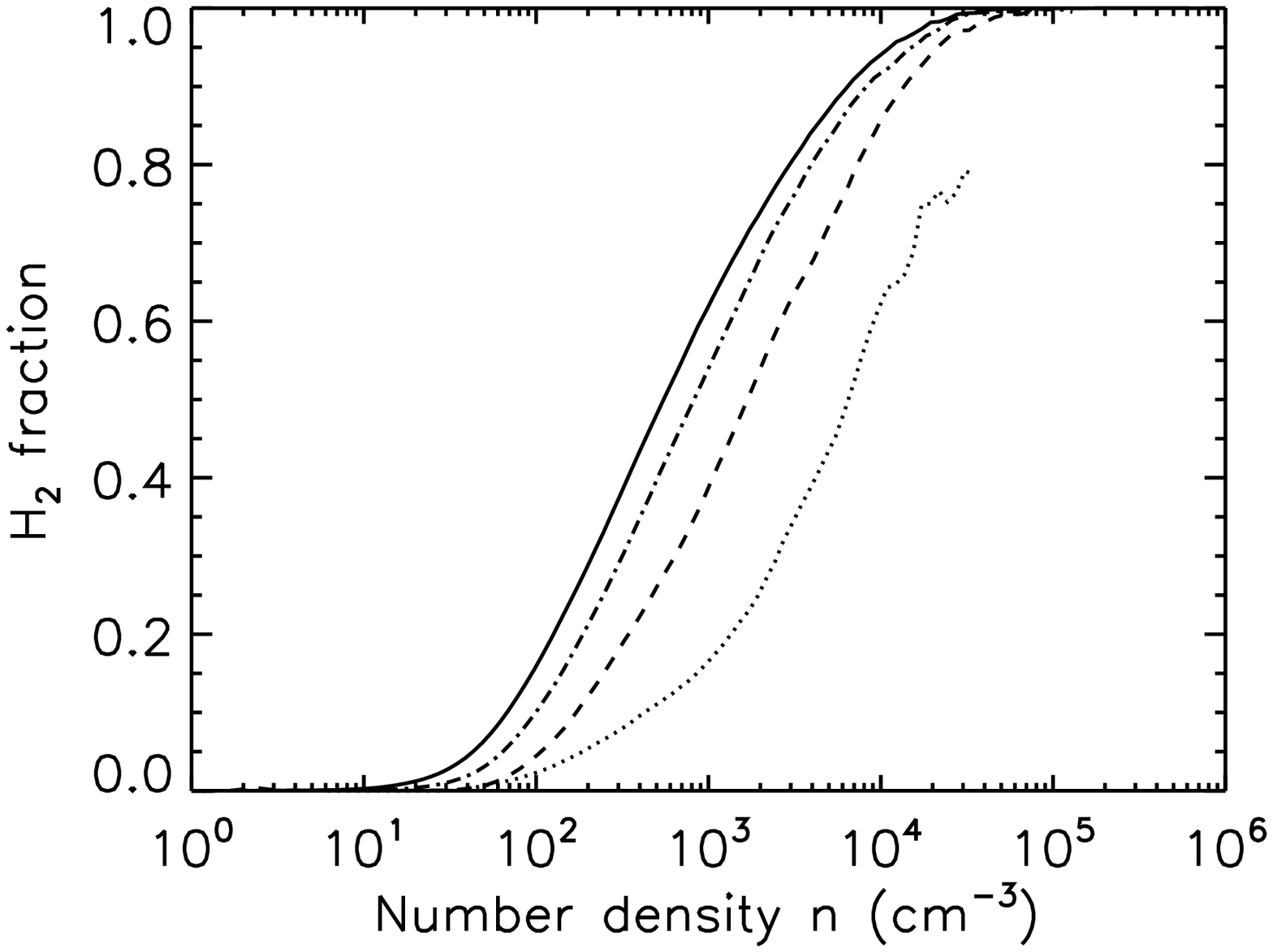,width=25pc,angle=0,clip=}
\epsfig{figure=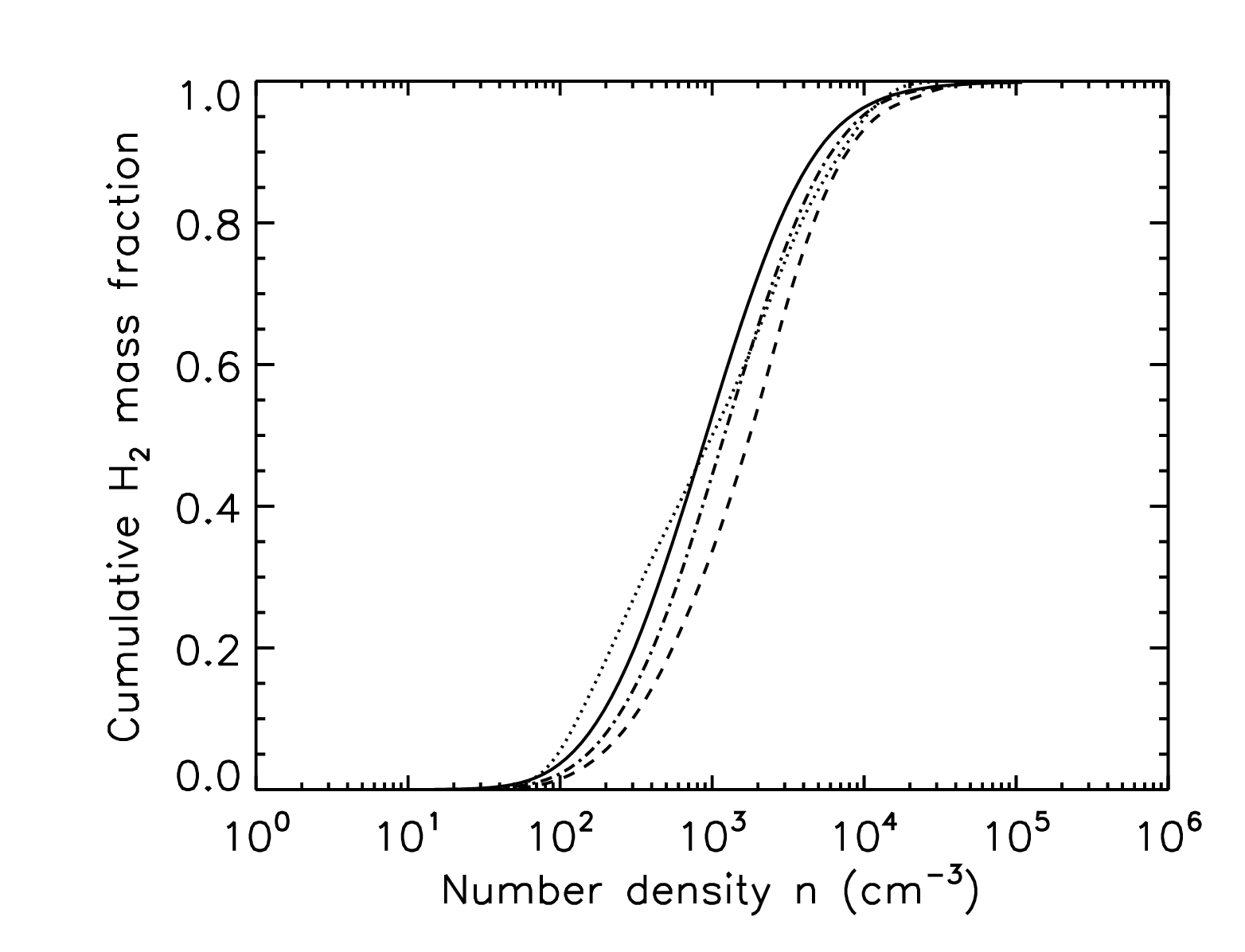,width=25pc,angle=0,clip=}
\figcaption{(a) $\mHt$ fraction as a function of number density $n$ in $512^{3}$ zone 
run MT512 at four different output times: $t = 0.6 \: {\rm Myr}$ (dotted line), 
$1.3 \: {\rm Myr}$ (dashed line), $1.9 \: {\rm Myr}$ (dot-dashed line) and
$2.5 \: {\rm Myr}$ (solid line). 
(b) As (a), but for  the cumulative mass distribution of $\mHt$ with $n$. 
\label{xh2vn-MT-times}}
\end{figure}

In Figure~\ref{xh2vn-MT-times}a, we show how the dependence of $x_{\mHt}$ on $n$ varies
as the gas evolves. For clarity, we have omitted any indication in this plot of the scatter 
around the mean values at the different output times, but in each case this is similar to the
scatter already seen in Figure~\ref{xh2vn-MT}. At $t = 0.6 \: {\rm Myr}$, most of the gas
is still atomic: the molecular fraction exceeds 50\% only at densities $n > 6000 \: 
{\rm cm^{-3}}$, and at this point in the simulation very little gas is found at these 
densities. Furthermore, a significant quantity of atomic hydrogen ($\sim 20\%$)
remains in this highly dense gas. By $t = 1.3 \: {\rm Myr}$, much more $\mHt$ has formed,
as is evident both from Figure~\ref{xh2vn-MT-times}a, and from our previous plot of the
time evolution of $\Htmass$ (Figure~\ref{H2-MT-fid}), which shows that it has increased from 
$\Htmass \simeq 0.07$ at $t = 0.6 \: {\rm Myr}$ to $\Htmass \simeq 0.25$ at 
$t = 1.3 \: {\rm Myr}$, an increase of almost a factor of four. We see from 
Figure~\ref{xh2vn-MT-times}a that gas at densities $n > 10^{3} \: {\rm cm^{-3}}$ is
now more than 50\% molecular, and that gas at densities $n > 10^{4} \: {\rm cm^{-3}}$
is almost entirely molecular. On the other hand, gas at densities $n < n_{\rm i}$ remains
almost entirely atomic. Between these limits, there seems to be a simple relationship
between the mean $x_{\mHt}$ fraction and the density, although this relationship is
not well described by a simple power-law fit. At later output times, the dependence of 
$x_{\mHt}$ on $n$ remains qualitatively similar to the relationship found at 
$t = 1.3 \: {\rm Myr}$, but the curve shifts to progressively lower densities, 
indicating that as $t$ increases, the mean $x_{\mHt}$ fraction found at a given gas
density also increases.


We have also computed the cumulative $\mHt$ mass distribution for run MT512 for the 
same set of output times. This is plotted in Figure~\ref{xh2vn-MT-times}b. We
see that at all of the output times that we examine, the majority of the $\mHt$
is found in the density range $10^{2} \leq n \leq 10^{4} \: {\rm cm^{-3}}$. As we
saw in the previous subsection, only a small fraction of the gas in our simulations 
has densities $n > 10^{4} \: {\rm cm^{-3}}$ and so little of the total $\mHt$ mass
is located at these densities. On the other hand, while there is a significant
amount of gas which has $n < 100 \: {\rm cm^{-3}}$, the molecular fraction in this
gas is small, and so its contribution to the total $\mHt$ mass is also small. 
Figure~\ref{xh2vn-MT-times}b suggests that lower density gas contributes a 
larger fraction of the total $\mHt$ mass at later times, consistent with the 
increase in the mean molecular fraction at fixed $n$ noted above: gas closer to 
the peak of the density PDF is becoming more molecular and so is contributing a 
larger fraction of the mass.

Finally, the fact that it takes $\sim 1 \: {\rm Myr}$ before the RMS density 
in our highest resolution runs stops increasing rapidly, as can be seen in 
Figure~\ref{dens-MT}, suggests that it takes at least this long for the 
turbulence to become fully developed. This implies that our results at times
$t < 1 \: {\rm Myr}$ will be significantly affected by initial transients, 
and so may be unrepresentative of the behaviour at later times. The results 
presented in Figure~\ref{xh2vn-MT-times} certainly appear to be consistent
with this conclusion, as the distribution of $\mHt$ at $t = 0.6 \: {\rm Myr}$ 
is qualitatively different from the distribution at our three later output 
times. This suggests that were we to allow the turbulence to develop fully in our 
simulations before beginning to follow the $\mHt$ formation, we would find 
somewhat different results at early times. Specifically, we would expect that
in this case the $\mHt$ formation timescale would be even shorter than that
found in the present work. However, it must be stressed that although our 
current set of initial conditions are rather artificial, deliberately suppressing 
$\mHt$ formation until after large density inhomogeneities have developed is equally 
artificial. Since our main aim in this paper is to put an upper limit on the 
$\mHt$ formation timescale, we believe that our current procedure is justified.

\clearpage

\section{Thermodynamics}
\label{thermo}
\subsection{Gas temperature: evolution and distribution}
\label{temp}
\begin{figure}
\centering
\epsfig{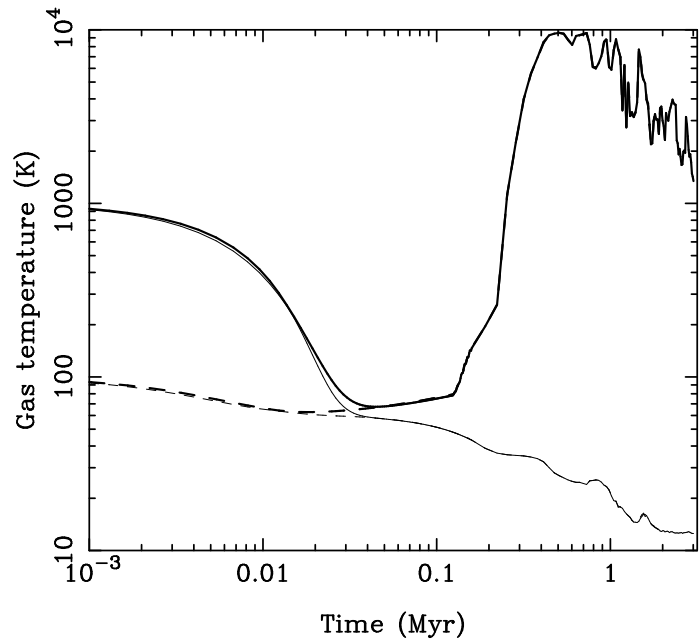}
\caption{Evolution of the maximum temperature of the gas, $T_{\rm max}$,
and the minimum temperature, $T_{\rm min}$ in $256^{3}$ zone runs MT256
(thick and thin solid lines) and MT256-T100 (thick and thin dashed lines).
In run MT256, the initial gas temperature $T_{\rm i} = 1000 \: {\rm K}$,
while in run MT256-T100, $T_{\rm i} = 100 \: {\rm K}$. \label{T-MT}}
\end{figure}

In order to verify that our results are not sensitive to the initial temperature
of the gas in our simulations, we ran a $256^{3}$ simulation (run MT256-T100) 
in which we set $T_{\rm i} = 100 {\rm K}$ but kept all of the other input 
parameters the same as in run MT256. The evolution with time of the minimum and 
maximum gas temperatures, $T_{\rm min}$ and $T_{\rm max}$, in this run and in run 
MT256 are plotted in Figure~\ref{T-MT}. We see that 
following a short initial period of cooling, the behaviour of the two runs is
indistinguishable. We have also verified that there is no significant difference
in the evolution of $\Htmass$ between the two runs.

\begin{figure}
\centering
\epsfig{figure=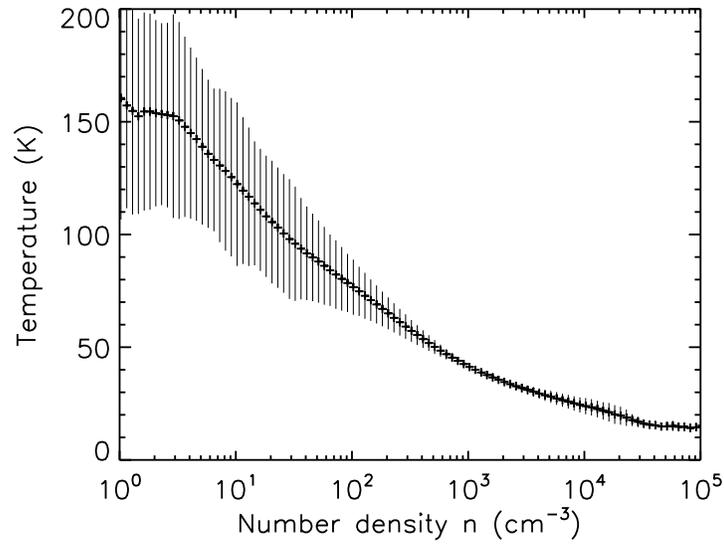,width=25pc,angle=0,clip=}
\caption{Mean gas temperature $T$ plotted as a function of the number density
$n$ in $512^{3}$ zone run MT512 at time $t = 1.9 \: {\rm Myr}$. The data were binned in a similar 
fashion to that used in the construction of Figure~\ref{xh2vn-MT} above. The standard 
deviation in the mean value in each bin is indicated wherever it exceeds the symbol 
size.}
\label{Tvn-MT}
\end{figure}

It is also interesting to examine how the gas temperature varies as a function of 
number density in these simulations. To do this, we took the output data from
run MT512, binned it by number density $n$, and then computed the mean temperature
and the standard deviation in the mean for each bin. The resulting values are shown in
Figure~\ref{Tvn-MT}.  Comparison of these results with those obtained from run MS256
in paper I demonstrates that we recover the same mean temperature in our fiducial 
turbulent run as in the analogous, initially static runs, at least within the range of 
number densities occupied by gas in the latter. This is unsurprising, as the short 
cooling times lead us to expect that most of the gas in both sets of simulations 
should be in thermal equilibrium, but it is good to verify that this is actually the 
case.

However, the results from our turbulent runs do differ from those from the initially static runs
of paper I in two important respects. The first is that gas is found at a very much wider range 
of densities in the turbulent runs, thanks to the strong compressions and rarefactions produced
by the turbulence. The second is that at low densities ($n \simless 10^{2} \: {\rm cm^{-3}}$), 
there is considerable scatter in the temperatures found in the gas, as can be seen from the
size of the error bars in Figure~\ref{Tvn-MT}; this scatter is absent in the static runs. The cause
of this scatter is the large number of strong shocks present in the turbulent simulations, or, 
more precisely, the large post-shock temperatures, which can be as large as a few thousand 
Kelvin, as can be seen from Figure~\ref{T-MT}. However, since the post-shock cooling 
regions are generally under-resolved in our simulations, it is likely that this scatter is 
overstated. 

\subsection{Effective polytropic index}
\label{gamma}
Another thermodynamical quantity of interest is the effective polytropic index of the
gas, $\gamma_{\rm eff}$, defined here as
\begin{equation}
\gamma_{\rm eff} = \frac{{\rm d} \ln p}{{\rm d} \ln \rho}, \label{geff}
\end{equation}
where $p$ is the gas pressure. Most simulations of turbulence within molecular 
clouds assume an isothermal equation of state, in which case $\gamma_{\rm eff} = 1$
at all densities. However, simulations of turbulent fragmentation performed by 
\citet{lkm03} for gas with a polytropic equation of state (in which case 
$\gamma_{\rm eff} \neq 1$ but is independent of $\rho$) and by \citet{japp05} for
gas with a piecewise polytropic equation of state ($p \propto \rho^{\gamma_{1}}$ for 
$\rho < \rho_{\rm c}$ and $p \propto \rho^{\gamma_{2}}$ for $\rho > \rho_{\rm c}$,
with $\gamma_{1} \neq \gamma_{2}$) show that the outcome of the fragmentation
process is dependent on the value of $\gamma_{\rm eff}$. Gas with a soft equation
of state (low $\gamma_{\rm eff}$) fragments more easily than gas with a hard 
equation of state (high $\gamma_{\rm eff}$), producing a larger number of fragments
with a smaller characteristic mass. In view of this, it is interesting to examine how
$\gamma_{\rm eff}$ varies with density in our simulations.

\begin{figure}
\centering
\epsfig{figure=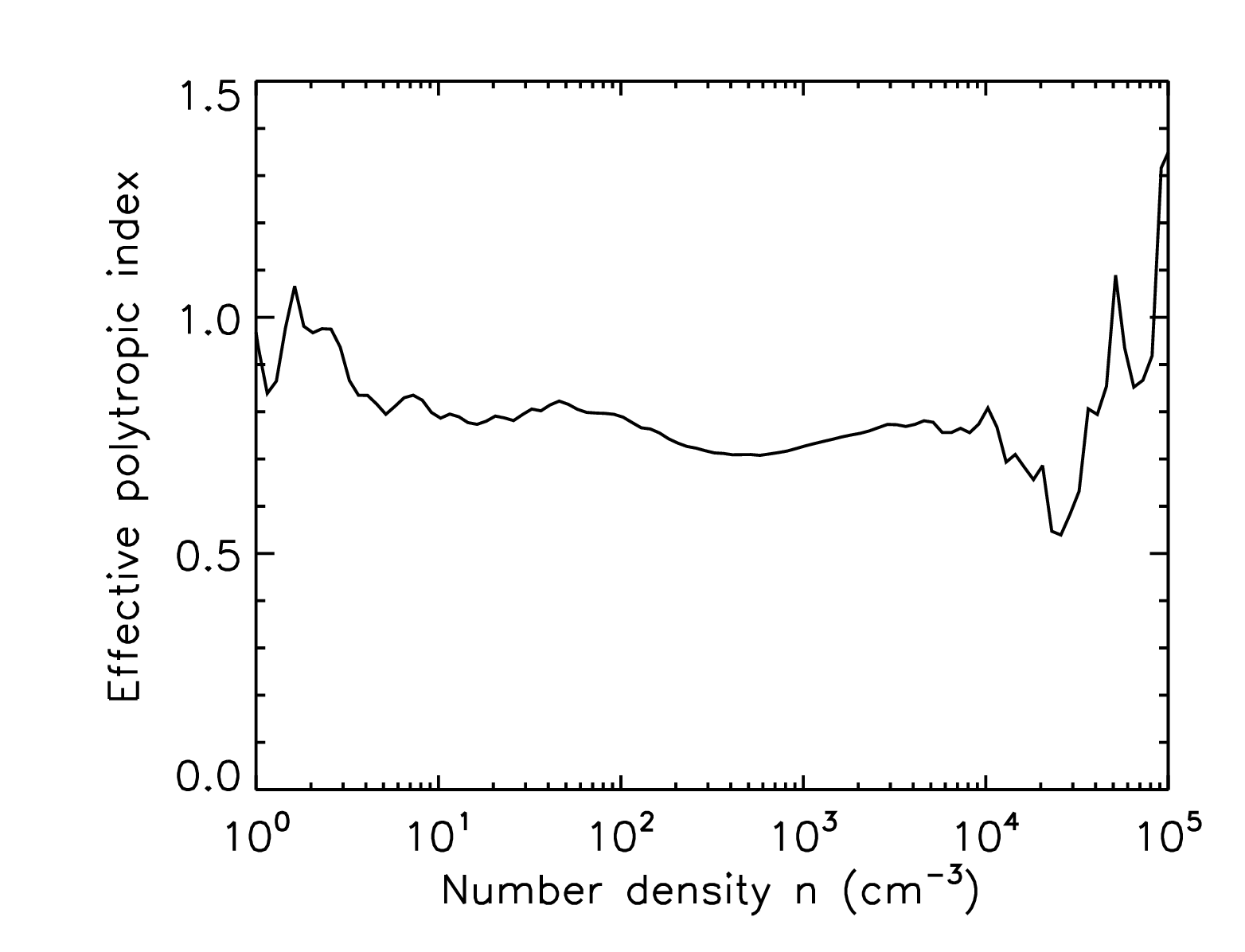,width=25pc,angle=0,clip=}
\caption{Value of $\gamma_{\rm eff}$ as a function of $n$ in 
$512^{3}$ zone run MT512 at time $t = 1.9 \: {\rm Myr}$. The data were binned as 
indicated in the text.}
\label{gamma-MT}
\end{figure}

The variation in $\gamma_{\rm eff}$ with $n$ is plotted in Figure~\ref{gamma-MT}
for run MT512 at $t = 1.9 \: {\rm Myr}$. To compute the values of $\gamma_{\rm eff}$ 
plotted here, we first computed the mean gas temperature as a function of density,
as in the previous section, and also the mean molecular fraction as a function of 
density, as in \S~\ref{morph}. We then used these values to compute the gas 
pressure as a function of density, and from this computed $\gamma_{\rm eff}$ through 
the use of equation~\ref{geff}. We see that although the gas in our simulations is not
a  simple polytrope, since $\gamma_{\rm eff}$ varies with density, its rate of 
change is small for densities in the range $10 < n < 10^{4} \: {\rm cm^{-3}}$,
suggesting that at these densities a polytropic approximation with 
$\gamma_{\rm eff} \simeq 0.8$ may actually be adequate for many applications.
This value of $\gamma_{\rm eff}$ is in reasonable agreement with the value of 
$\gamma_{\rm eff} = 0.725$ derived by \citet{japp05} for these densities from a 
synthesis of various observational and theoretical sources. However, it is significantly 
smaller than the values computed by \citet{ss00} for this range of densities. 

At densities $n < 10 \: {\rm cm^{-3}}$, we find larger values
of $\gamma_{\rm eff}$, indicating a stiffening of the equation of state. However,
at these densities the large scatter present in the gas temperature implies that
the gas is not well described by an equation of state that is purely a function
of density and so the values of $\gamma_{\rm eff}$ that we have derived for
this density regime are of questionable validity. 

We also discount the large fluctuations we see in $\gamma_{\rm eff}$ at high 
densities ($n > 10^{4} \: {\rm cm^{-3}}$). These are caused by small fluctuations 
in the mean temperature, with bin-to-bin variations of less than a degree, and
our treatment of cooling at these densities is not sufficiently accurate for us to 
trust in the reality of these features.

\subsection{Influence of the cosmic ray ionization rate}
\label{cr_sec}
As previously discussed in paper I, measurements of the cosmic ray ionization
rate in diffuse gas \citep{mac03,liszt03,lrh04} and dense gas 
\citep{cwth98,bpwm99,vv00} give results that differ by an order
of magnitude or more. In most of our simulations, both in paper I and here,
we adopted an ionization rate $\zeta = 10^{-17} \: {\rm s^{-1}}$, consistent 
with the value measured in dense gas. However, in order to establish the 
sensitivity of our results to the assumed ionization rate, we also
performed one $256^{3}$ zone simulation with 
$\zeta = 10^{-15} \: {\rm s^{-1}}$, a value that lies near the high end of current
determinations. Aside from $\zeta$, the remaining input parameters of this
simulation, which we designate MT256-CR, were the same as in run MT256, and
so comparison of these two runs serves to highlight the effects of varying 
$\zeta$.

\begin{figure}
\centering
\epsfig{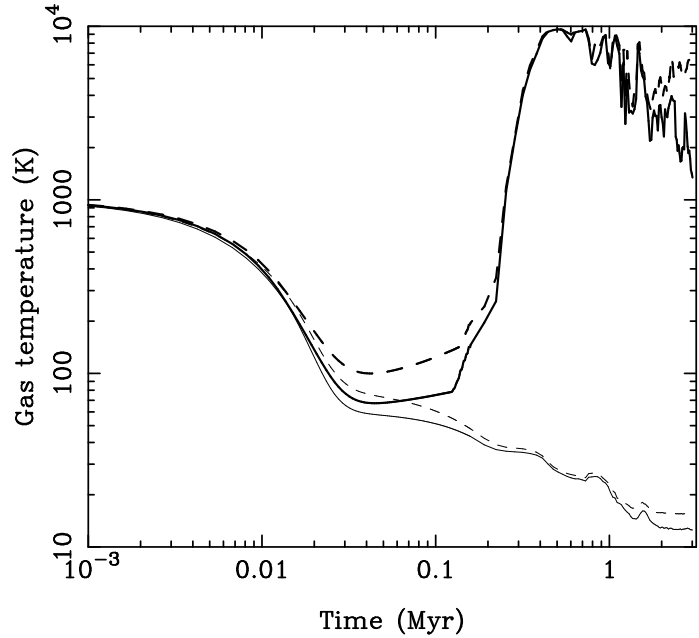}
\caption{Evolution of the maximum temperature of the gas, $T_{\rm max}$,
and the minimum temperature, $T_{\rm min}$ in $256^{3}$ zone runs MT256
(thick and thin solid lines) and MT256-CR (thick and thin dashed lines),
which were performed with cosmic ray ionization rates 
$\zeta = 10^{-17} \: {\rm s^{-1}}$ and $\zeta = 10^{-15} \: {\rm s^{-1}}$ respectively.
\label{T-MT-CR}}
\end{figure}

We compare the evolution of $T_{\rm min}$ and $T_{\rm max}$ in runs MT256 and
MT256-CR in Figure~\ref{T-MT-CR}. We see two main effects: both $T_{\rm min}$
and $T_{\rm max}$ are increased by about 10--20~K at times 
$0.02 < t < 0.2 \: {\rm Myr}$, and $T_{\rm max}$ (but not $T_{\rm min}$) is
increased by a factor of a few at times $t > 2\: {\rm Myr}$. Interestingly, 
there is little difference between the runs in the interval 
$0.2 < t < 2 \: {\rm Myr}$ during which much of the $\mHt$ forms. 
We would therefore expect to find little difference in the $\mHt$ formation
rates in the two simulations, and this is borne out by our results. After
$2 \: {\rm Myr}$ of evolution, run MT256-CR has formed only 2\% more $\mHt$
than run MT256, demonstrating that the influence of variations in $\zeta$ is
small.

\section{Sensitivity to variations of the input parameters}
\label{sense}
As with the runs discussed in paper I, it is interesting
to ask what happens as we vary some of the more important input parameters.
The effects of varying the initial temperature of the gas have already been
considered in the previous section; the effects of varying the box size $L$,
the initial magnetic field strength $B_{\rm i}$, the initial turbulent velocity dispersion 
$v_{\rm rms, i}$ and the initial density $n_{\rm i}$ are examined
in \S~\ref{turb_box}--\ref{turb_n0} below.

\subsection{Box size}
\label{turb_box}
\begin{figure}
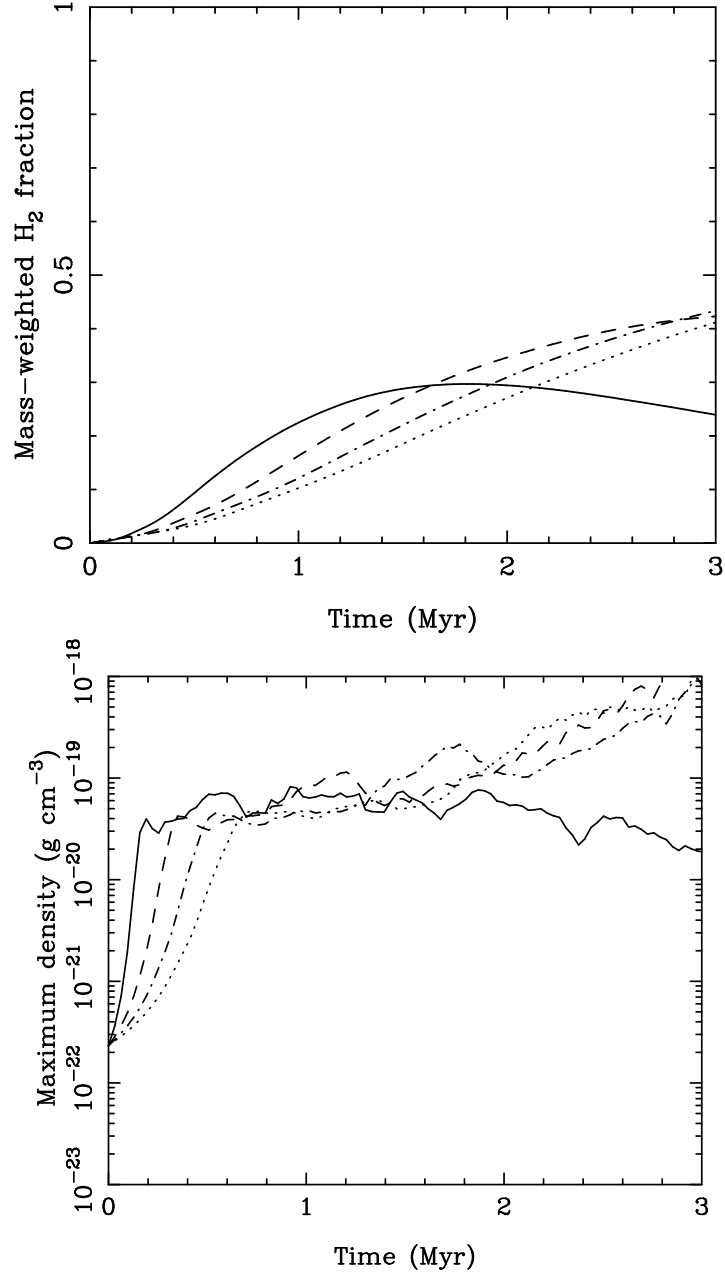

\centering
\epsfig{figure=fig20a.eps,width=20pc,angle=270,clip=}
\epsfig{figure=fig20b.eps,width=20pc,angle=270,clip=}
\caption{(a) Evolution of $\Htmass$ with time in $256^{3}$ zone
runs MT256-L10 (solid line), MT256 (dashed line), MT256-L30 
(dot-dashed line) and MT256-L40 (dotted line), which had box sizes
$L = 10$, 20, 30 and $40 \: {\rm pc}$ respectively.
(b) As (a), but for the time evolution of $\rho_{\rm max}$.
\label{H2-MT-box}}
\end{figure}

To investigate the extent to which the quantity of $\mHt$ formed in our simulations
and the rate at which it forms depend upon our choice of box size $L$, we have 
performed further $256^{3}$ zone simulations with $L = 10$, 30 and $40 \: {\rm pc}$, 
which we have designated as runs MT256-L10, MT256-L30 and MT256-L40 respectively.
The evolution of $\Htmass$ in these simulations is plotted in Figure~\ref{H2-MT-box}a, 
along with the results from run MT256 for comparison. We see that as we decrease $L$, 
we increase the amount of $\mHt$ to form within the first 1--2~Myr of the simulation. 
This is most likely due to the fact that the network of dense sheets and filaments
that characterizes the density field in our turbulent runs takes less time to develop
in simulations run with a smaller $L$, since the turbulent crossing time of the box
is directly proportional to $L$. Evidence for this can be seen in our plot of 
$\rho_{\rm max}$ versus time in Figure~\ref{H2-MT-box}b. 

In runs MT256, MT256-L30 and MT256-L40 the differences resulting from the change in
$L$ are small, and have mostly vanished by $t = 3 \: {\rm Myr}$, with the value of
$\Htmass$ in all three runs having converged at about 0.4. The behaviour of run 
MT256-L10, on the other hand, differs markedly: $\Htmass$ reaches a peak value at
$t \sim 1.6 \: {\rm Myr}$ of $\Htmass \sim 0.3$ and then begins to decline, decreasing
by approximately 20\% by $t = 3 \: {\rm Myr}$. The reason for this fall off in 
$\Htmass$ is hinted at in Figure~\ref{H2-MT-box}b: in the $L = 10 \: {\rm pc}$ 
simulation, none of the gas in the dense shocked regions undergoes gravitational 
collapse, and so the peak density begins to decrease as the turbulence decays. The 
total amount of dense gas also decreases as the turbulence decays, and so increasing
amounts of $\mHt$ formed in high density gas finds itself in moderate or low density 
regions where $\mHt$ photodissociation occurs faster than $\mHt$ formation. Consequently, the
total amount of $\mHt$ present in the simulation declines. However, it should be 
noted that since our treatment of $\mHt$ photodissociation leads us to overestimate 
the $\mHt$ photodissociation rate, it also causes this decline to occur much faster
in our simulations than would be the case in a real cloud. 

\subsection{Initial magnetic field strength}
\label{mag_field}
We ran three $256^{3}$ simulations in which we varied only $B_{\rm i}$ in order to
determine how sensitive our results are to the strength of the magnetic field. In the
first of the simulations, run HT256, we set $B_{\rm i} = 0.0$ and so examined purely
hydrodynamical turbulence. In the second, run MT256-Bx2, we set $B_{\rm i} = 
11.7 \mu {\rm G}$, twice our standard value, while in the third, run MT256-Bx10, we
set $B_{\rm i} = 58.5 \mu {\rm G}$, ten times as large as our standard value. The 
mass-to-flux ratios in these three runs are $M / \Phi = \infty, 2, 0.4$ respectively; note 
that the gas in run MT256-Bx10 is magnetically subcritical and so will not undergo
gravitational collapse. 

\begin{figure}
\centering
\epsfig{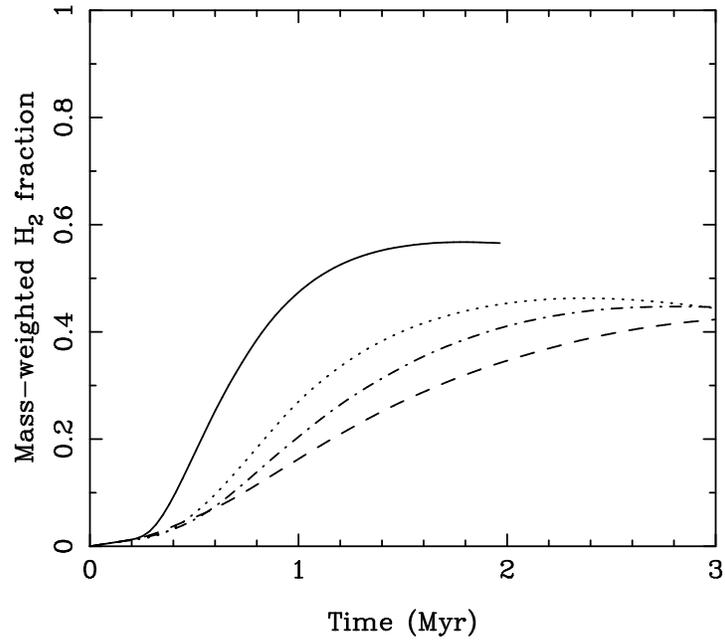}
\caption{Evolution of $\Htmass$ with time in $256^{3}$ zone runs with different
magnetic field strengths. We plot results for runs HT256 (solid line), MT256 (dashed line), 
MT256-Bx2 (dot-dashed line) and MT256-Bx10 (dotted line), which have initial 
field strengths $B_{\rm i} = 0.0$, 5.85, 11.7 \& $58.5 \:  \mu {\rm G}$ respectively.}
\label{H2-MT-mag-comp}
\end{figure}

\begin{figure}
\centering
\epsfig{figure=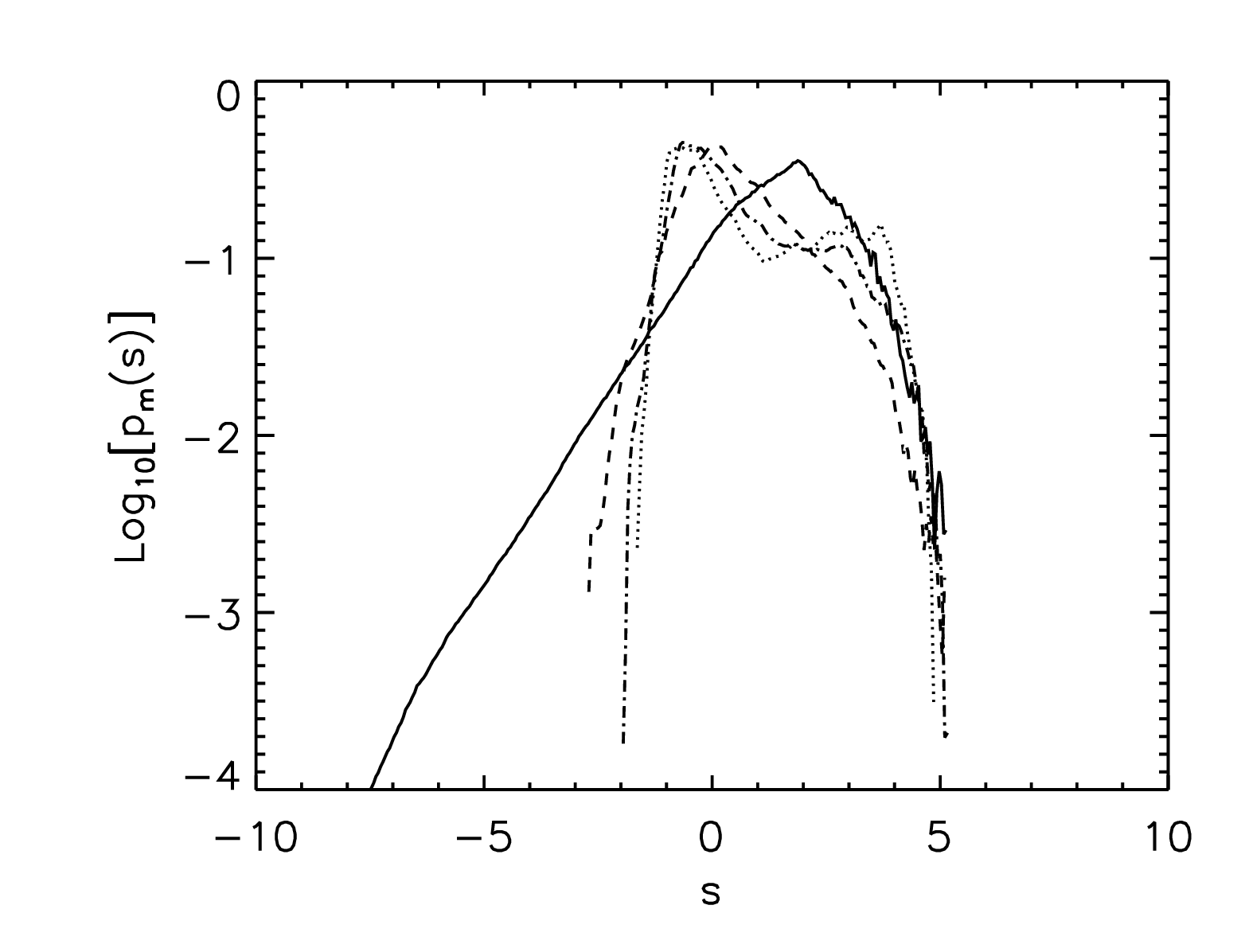,width=25pc,angle=0,clip=}
\epsfig{figure=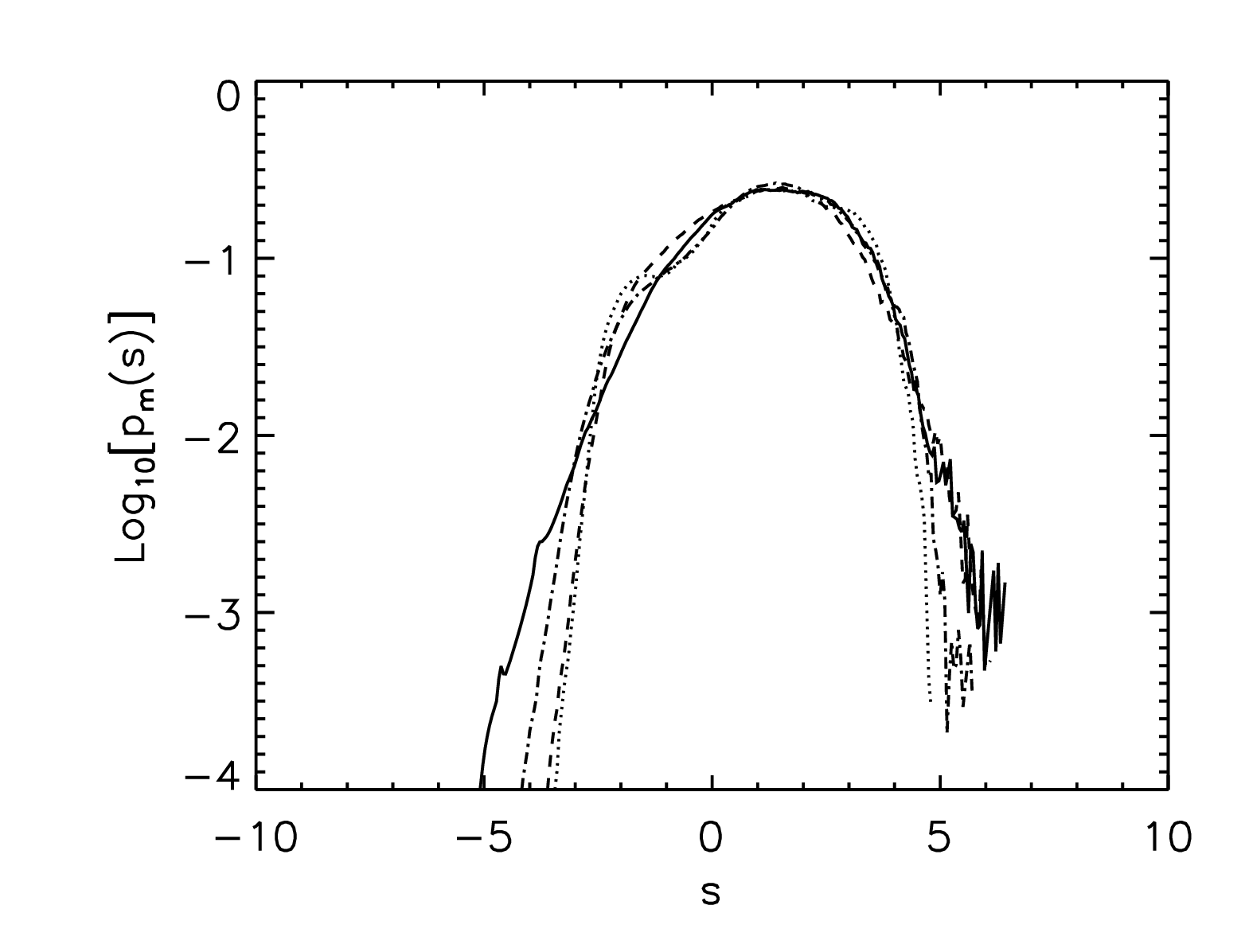,width=25pc,angle=0,clip=}
\caption{(a) Mass-weighted density PDF at $t = 0.63 \: {\rm Myr}$ in 
$256^{3}$ zone runs HT256 (solid line), MT256 (dashed line), 
MT256-Bx2 (dot-dashed line) and MT256-Bx10 (dotted line), 
performed with different initial magnetic field strengths.
(b) As (a), but for $t = 1.9 \: {\rm Myr}$.  \label{MT-mag-pdf}}
\end{figure}

\begin{figure}
\centering
\epsfig{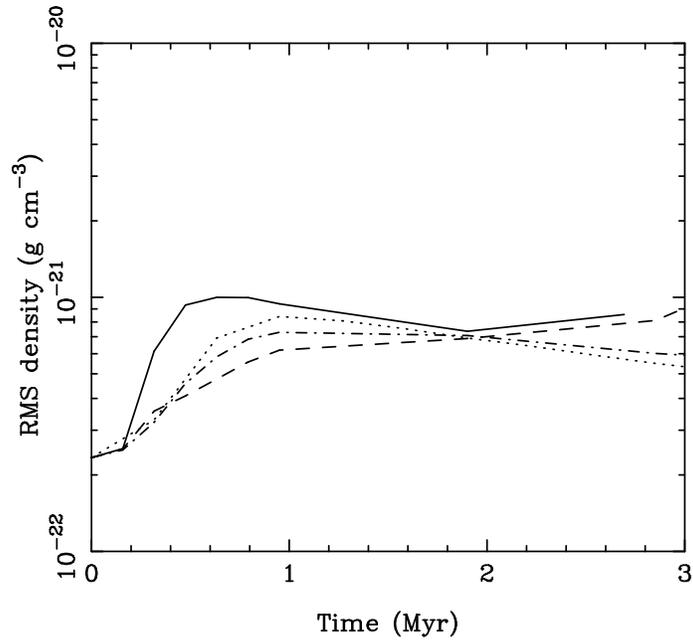}
\caption{Evolution of $\rho_{\rm rms}$ with time in $256^{3}$ zone
runs HT256 (solid line), MT256 (dashed line), MT256-Bx2 (dot-dashed line) 
and MT256-Bx10 (dotted line), performed with different initial magnetic field 
strengths.  \label{MT-mag-drms}}
\end{figure}

In Figure~\ref{H2-MT-mag-comp}, we show how $\Htmass$ evolves in these three runs
and how this compares to its evolution in run MT256. Clearly, the growth of $\Htmass$ 
is more rapid in the $B_{\rm i} = 0.0$ run than in any of the magnetized runs. This is not
unexpected: in our $B_{\rm i} \neq 0.0$ runs, the magnetic pressure of the gas helps it
to resist compression by the turbulence, and so we would expect it to take longer to build
up the same amount of high density structure in the $B_{\rm i} \neq 0.0$ runs as in the
$B_{\rm i} = 0.0$ run. Indeed, if we compare how the density PDFs of the four runs evolve,
we find that at early times, there is significantly more dense gas in run HT256 than in any
of the magnetised runs, as illustrated in Figure~\ref{MT-mag-pdf}a, while by 
$t \sim 2 \: {\rm Myr}$ the PDFs in the four runs have almost converged, at least for 
densities near the peak (Figure~\ref{MT-mag-pdf}b). Similarly, if we compare the evolution 
of $\rho_{\rm rms}$ in the four runs (Figure~\ref{MT-mag-drms}), we see that at 
$t < 1 \: {\rm Myr}$, $\rho_{\rm rms}$ is between 50\% and 100\% larger in run MT256 than 
in any of the magnetized runs.

\begin{figure}
\centering
\epsfig{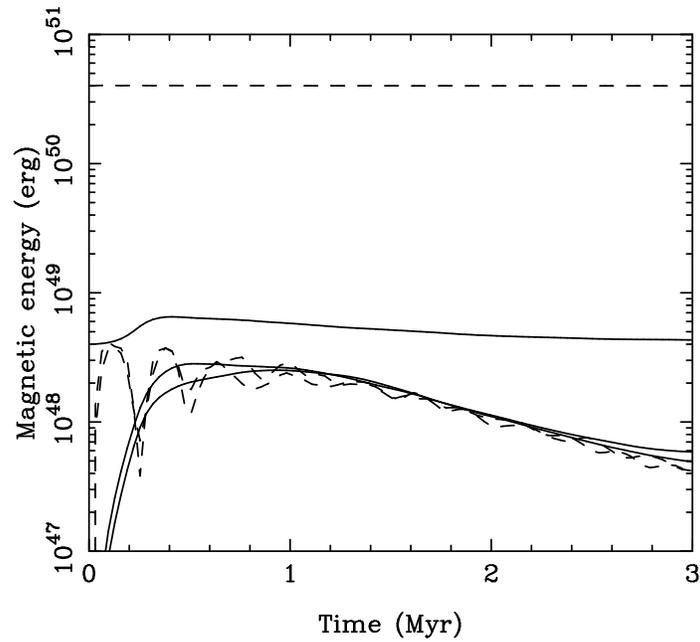}
\caption{Evolution with time of the $x$, $y$ and $z$ components of the
magnetic energy in $256^{3}$ runs MT256 (solid lines) and MT256-Bx10 
(dashed lines). The initial magnetic field strength in run MT256-Bx10 is
ten times stronger than in run MT256. In both cases, the uppermost line 
corresponds to the $z$ component of the field.  \label{MT-mag-en}}
\end{figure}

Another interesting feature of Figure~\ref{H2-MT-mag-comp} is the fact that in the 
$B_{\rm i} \neq 0.0$ runs, the relationship between $\Htmass$ and $B_{\rm i}$ is
not what we might expect: as we increase $B_{\rm i}$, and so increase the 
magnetic pressure of the gas, we find that $\Htmass$ also increases. This behaviour
is due to an effect previously noted by \citet{hmk01} in their simulations of
isothermal MHD turbulence. They note that as the strength of the magnetic field
increases, the density enhancements found in their simulations become greater. 
They attribute this to the fact that when the magnetic field is strong, it remains
highly ordered, causing the magnetic pressure to remain highly anisotropic.
Gas can therefore
flow along the field lines, forming dense shocked layers that are oriented perpendicularly 
to the mean field. On the other hand, when the field is weak it is easily tangled by
the turbulent velocity field, and so the magnetic pressure becomes far more isotropic,
making it harder to form highly dense regions. Figures~\ref{MT-mag-pdf} and 
\ref{MT-mag-drms}, which show that more dense gas is present at early times in runs
MT256-Bx2 and MT256-Bx10 than in run MT256, suggest that a similar effect is at work
in our simulations. Further confirmation of this comes from an examination of the
magnetic energy associated with the x, y and z components of the magnetic field in 
runs MT256 and MT256-Bx10, as plotted in Figure~\ref{MT-mag-en}. In run MT256-Bx10, 
the magnetic energy associated with the z-component of the field remains orders of 
magnitude larger than the energy associated with either the x or y-components throughout 
the simulation, demonstrating that the field remains highly anisotropic. In run MT256, 
on the other hand, the difference between the three components is much smaller, indicating 
that the magnetic field is far more isotropic.

Finally, the significant $\mHt$ formation that occurs in magnetically subcritical run
MT256-Bx10, which is not gravitationally unstable, is further confirmation that 
gravitational instability does not cause the rapid $\mHt$ formation seen in our 
simulations, and that turbulent compressions dominate even in strongly magnetized
gas. 

\subsection{Initial velocity dispersion}
\label{turb_vrms}
\begin{figure}
\centering
\epsfig{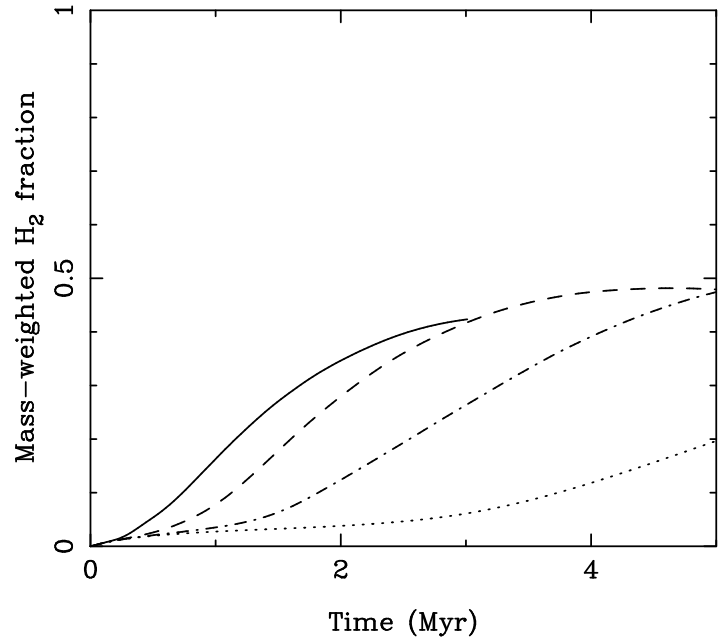}
\caption{Evolution of $\Htmass$ with time in $256^{3}$ zone runs with 
different initial RMS turbulent velocities. We plot results for runs MT256 
(solid line), MT256-v5 (dashed line), MT256-v2.5 (dot-dashed line) and 
MT256-v1 (dotted line), which had $v_{\rm rms, i} = 10$, 5, 2.5 \& 
$1 \: {\rm km} \: {\rm s^{-1}}$ respectively. \label{H2-MT-vdisp-comp}}
\end{figure}

To examine the extent to which our results depend upon the strength of the turbulence,
as quantified by the initial rms velocity, $v_{\rm rms, i}$, we have run several simulations
with differing values of $v_{\rm rms, i}$. In Figure~\ref{H2-MT-vdisp-comp}, we show how
$\Htmass$ evolves in runs MT256-v1, MT256-v2.5 and MT256-v5, which have 
$v_{\rm rms, i} = 1.0$, 2.5 and $5.0 \: {\rm km} \: {\rm s}^{-1}$ respectively. We see that
as we decrease the strength of the turbulence, we increase the $\mHt$ formation 
timescale, with the time taken to reach $\Htmass \sim 0.4$ approximately doubling from
$t \sim 2 \: {\rm Myr}$ to $t \sim 4 \: {\rm Myr}$ as we decrease $v_{\rm rms, i}$ from
$10 \: {\rm km} \: {\rm s^{-1}}$ to $2.5 \: {\rm km} \: {\rm s^{-1}}$. However, the impact of
varying $v_{\rm rms, i}$ on the {\em amount} of $\mHt$ to form during the simulation is
surprisingly small: runs MT256, MT256-v5 and MT256-v2.5 have all formed roughly the
same amount of $\mHt$ by the time the simulations end, and only in run MT256-v1
is the final value of $\Htmass$ significantly smaller.

\begin{figure}
\centering
\epsfig{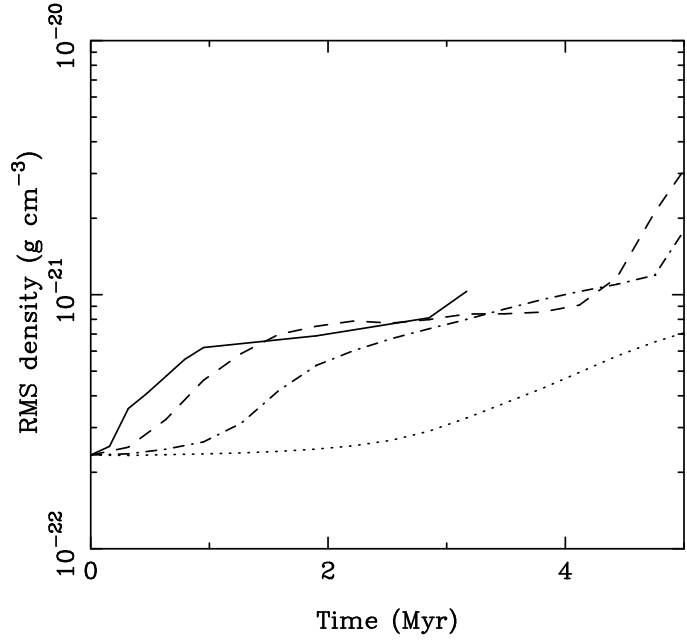}
\caption{Evolution of $\rho_{\rm rms}$ with time in $256^{3}$ zone runs MT256 (solid line),
MT256-v5 (dashed line), MT256-v2.5 (dot-dashed line) and MT256-v1 (dotted line),
performed with different initial RMS turbulent velocities. \label{drms-vrms}}
\end{figure}

The reason we find faster $\mHt$ formation in runs with a larger $v_{\rm rms, i}$ 
becomes clear if we examine how the RMS density of the gas varies in these runs. In 
Figure~\ref{drms-vrms}, we plot the time evolution of $\rho_{\rm rms}$ in all four runs.
We see that as we increase $v_{\rm rms, i}$, we find a faster increase in $\rho_{\rm rms}$
at early times. This behaviour is easily understood on the basis of a simple timescale 
argument: runs with a larger $v_{\rm rms, i}$  have a shorter turbulent crossing time,
$t_{\rm cross} = L / v_{\rm rms}$, and so in these runs less time is required to produce
the highly overdense regions where most of the $\mHt$ forms. At $t \sim 3 \: {\rm Myr}$
in run MT256 and at $t \sim 4.5 \: {\rm Myr}$ in runs MT256-v5 and MT256-v2.5, we see
a sudden increase in $\rho_{\rm rms}$, consistent with the onset of runaway gravitational
collapse at one or more locations in the simulation volume, but prior to this the 
evolution of the RMS density in these runs is dominated by turbulence. Since the turbulent
compressions produce a very similar RMS density in all three cases, it is not surprising
that we also find similar values for $\Htmass$. Note also that the spread in $\mHt$ formation 
timescales seen in Figure~\ref{H2-MT-vdisp-comp} is comparable to the spread in RMS
density growth rates in Figure~\ref{drms-vrms}, and that the values of $\Htmass$ in the 
three runs converge later than the values of $\rho_{\rm rms}$ because $\mHt$ formation is 
not instantaneous, causing the growth in $\Htmass$ to lag behind the growth in $\rho_{\rm rms}$.

Finally, although run MT256-v1 appears, on the basis of Figure~\ref{H2-MT-vdisp-comp}, to 
behave differently from the other runs, Figure~\ref{drms-vrms} suggests that this is not 
actually the case: it is simply taking significantly longer to produce large overdensities 
and the associated high $\mHt$ fractions in this run, which is to be expected given the
smaller $v_{\rm rms}$ and hence greater turbulent crossing time. 


\subsection{Initial density}
\label{turb_n0}
\begin{figure}
\centering
\epsfig{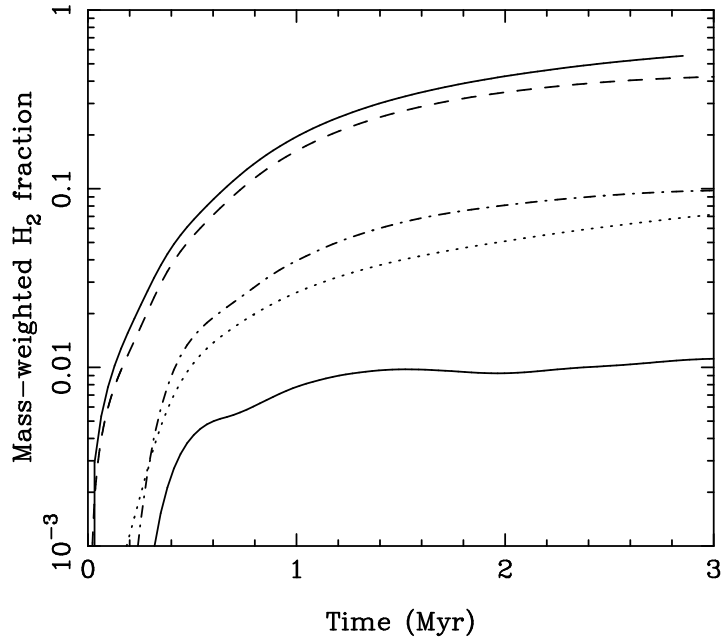}
\caption{Evolution of $\Htmass$ with time in $256^{3}$ zone
runs MT256-RT (upper solid line), MT256 (dashed line), 
MT256-n30 (dot-dashed line), MT256-RT-n10 (dotted line) and
MT256-n10 (lower solid line), performed with different initial 
gas densities. \label{H2-MT-n0}}
\end{figure}

In Figure~\ref{H2-MT-n0}, we show how the evolution with time of $\Htmass$ 
changes as we reduce the initial density of the gas in our simulations. 
Plotted in the figure are results from two runs performed using the local
shielding approximation, MT256-n10 and MT256-n30, which had initial densities 
of  $n_{\rm i} = 10$ and $30 \: {\rm cm^{-3}}$ respectively, and one run 
performed using the six-ray shielding approximation, MT256-RT-n10, which had
an initial density $n_{\rm i} = 10 \: {\rm cm^{-3}}$. To ensure that the
mass-to-flux ratio $M/\Phi$ remained approximately the same in these runs
as in our runs with $n_{\rm i} = 100 \: {\rm cm^{-3}}$, we reduced $B_{\rm i}$
to $1.755 \: \mu {\rm G}$ in run MT256-n30 and to $0.585 \: \mu {\rm G}$ 
in runs MT256-n10 and MT256-RT-n10. All of the other input parameters had the 
same values as in run MT256.

It is clear from Figure~\ref{H2-MT-n0} that reducing $n_{\rm i}$ has a dramatic
effect on the amount of $\mHt$ that forms in the simulations, particularly in 
the runs using the local shielding approximation. The decrease in $n_{\rm i}$
by about a factor of three between runs MT256 and MT256-n30 causes a roughly
proportionate decrease in $\Htmass$ at late times. Similarly, the reduction 
in $n_{\rm i}$ by an order of magnitude between runs MT256-RT and MT256-RT-n10
leads to an order of magnitude decrease in $\Htmass$. The difference between
runs MT256 and MT256-n10 is even more striking, with an order of magnitude
decrease in $n_{\rm i}$ leading in this case to a reduction in $\Htmass$ by
more than a factor of 40.

\begin{figure}
\centering
\epsfig{figure=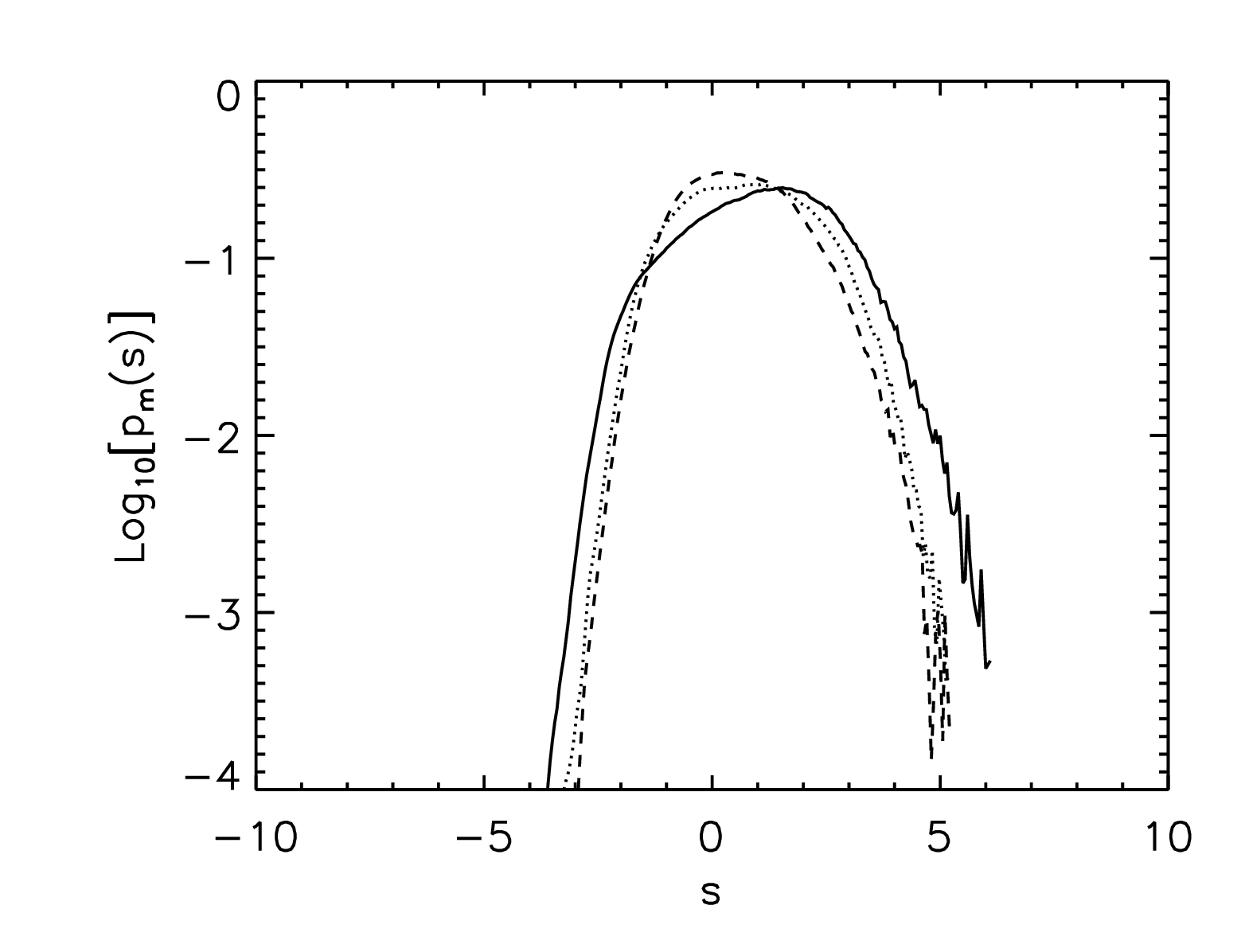,width=25pc,angle=0,clip=}
\caption{(a) Mass-weighted density PDF at $t = 1.9 \: {\rm Myr}$ in 
$256^{3}$ zone runs MT256 (solid line), MT256-n30 (dotted line) 
and MT256-n10 (dashed line), performed with different initial gas
densities. \label{pdf-MT-n0}}
\end{figure}

The reason for this sensitivity is clear if we look at how the density distribution
varies as we vary $n_{\rm i}$. In Figure~\ref{pdf-MT-n0}, we plot the mass-weighted 
density PDF for 
the suite of runs at different densities 
at $t = 1.9 \: {\rm Myr}$. 
In each case, $s = \ln(n / n_{\rm i})$, where the mean density $n_{\rm i}$ is of 
course different in each run. As we decrease $n_{\rm i}$, the peak of the PDF 
moves closer to the mean and the amount of overdense gas decreases. Since the 
mean density has also decreased, the net effect is a sharp drop in the amount 
of high density gas: in run MT256, at this output time, approximately 22\% of 
the total gas mass is in regions with  $n > 10^{3} \: {\rm cm^{-3}}$, while in 
runs MT256-n30 and MT256-n10, this figure decreases to 2\% and 0.05\%, respectively.
The density PDFs in 
the six-ray runs strongly resemble those in the local approximation
runs with the same initial density, 
and so a similar effect occurs in this case.

Since we already know that most of the $\mHt$ that forms in our 
simulations does so in dense gas, it is not surprising that 
substantially reducing the amount of dense gas available has a 
dramatic effect on $\Htmass$. It is clear from a comparison of 
runs MT256-n10 and MT256-RT-n10 in Figure~\ref{H2-MT-n0}, 
however, that the magnitude of this effect does depend on our
choice of self-shielding approximation: we see a significantly
greater reduction when using the local shielding approximation
than when using the six-ray approximation. This is unsurprising
given our previous discussion of the dependence of $x_{\mHt}$
on $n$ in our turbulent models (see \S~\ref{xh2n}). We have 
already seen that far more $\mHt$ survives at low densities 
when we use the six-ray approximation. This low density 
$\mHt$ represents only a small fraction of the total molecular 
mass present in our $n_{\rm i} = 100 \: {\rm cm^{-3}}$ simulations, 
but is far more significant in simulations with a lower value of 
$n_{\rm i}$. Nevertheless, even in the six-ray run there is a
large reduction in $\Htmass$ when $n_{\rm i}$ is reduced, and
so although the quantitative details of the relationship 
between $n_{\rm i}$ and $\Htmass$ depend on the treatment of 
self-shielding, the basic qualitative result 
does not: conversion of $\mH$ to $\mHt$ is strongly suppressed at 
$n \simless 10 \: {\rm cm^{-3}}$. 

\clearpage

\section{Summary}
\label{summary}
The simulations presented in this paper demonstrate that the timescale for $\mHt$ 
formation in turbulent gas is much shorter than the corresponding timescale in 
quiescent gas. For gas with a mean number density $n_{\rm i} = 100 \: {\rm cm^{-3}}$,
and with a magnetic field strength and RMS turbulent velocity consistent with 
observations, we find that $\sim 40\%$ of the initial atomic hydrogen can be 
converted to $\mHt$ on a timescale of 1--$2 \: {\rm Myr}$. Moreover, since 
these results are derived from simulations 
that
use a local approximation to
treat $\mHt$ shielding, they represent a {\em lower limit} on the amount of
$\mHt$ formed. 
(Comparison between a run using a
a six-ray 
shielding approximation and one 
with 
$\mHt$ photodissociation 
entirely absent
suggests that this lower limit must lie within about 30\% of the true value).
The $\mHt$ formation timescale that we obtain is consistent with
that required by models in which molecular clouds have short lifetimes, of the 
order of a turbulent crossing time (see e.g.\ \citealt{elm00}, \citealt{hbpb01}), 
and so criticism of these models on the grounds that they do not provide enough 
time to form the large quantities of $\mHt$ required is shown to be groundless.

By comparing the rate of $\mHt$ formation in simulations performed both with and 
without self-gravity, we have shown that the rapid growth of $\mHt$ is due to
the large density enhancements created by the turbulent compressions. Self-gravity
becomes important only at later times, as part of the gas goes into runaway 
gravitational collapse. We therefore predict that $\mHt$ should form rapidly in any
dense turbulent atomic cloud, regardless of whether or not the cloud is 
gravitationally bound and regardless of whether it is magnetic subcritical or 
supercritical. Our finding that $\mHt$ forms quickly in a supersonically turbulent
flow is consistent with the previous work of \citet{pav02}, who find a similar effect
in simulations of high-speed decaying turbulence in $n = 10^{6} \: {\rm cm^{-3}}$
gas.

We have investigated the distribution of the $\mHt$ formed in our simulations, 
and have shown that most is located at densities between 1 and 100 times our
mean number density $n_{\rm i} = 100 \: {\rm cm^{-3}}$. Gas denser than $10^{4} \: 
{\rm cm^{-3}}$ is fully molecular, but contributes little to the total $\mHt$ mass as
only a small fraction of the gas is found at these densities. On the other hand, 
gas that is less dense than $100 \: {\rm cm^{-3}}$ contributes little because it is
still primarily atomic. 

A surprising result of our simulations is that in low density regions ($n < 300 \: 
{\rm cm^{-3}}$), there is more $\mHt$ than we would expect if the gas were in
photodissociation equilibrium. We have shown that this is explained by the 
fact that a large fraction of the $\mHt$ found at low densities actually formed
at high densities, in gas with $n > 1000 \: {\rm cm^{-3}}$, and that it has
subsequently been transported to low densities by the action of the turbulence.
This is consistent with previous findings that a large proportion of the dense 
structures produced in a supersonically turbulent flow are transient objects
which are not gravitationally bound \citep{khm00,vsks05}. Our results show that
these transient density enhancements have a profound impact on the $\mHt$ 
chemistry of the low density gas. Previous work by \citet{gwhrv05,gwr06}, 
using a highly simplified dynamical model, suggests that such transient density 
enhancements will also have a large impact on the chemistry of many other
tracer species, such as CS. We might also speculate that strong, transient 
density enhancements associated with turbulence in the diffuse ISM are the 
cause of the puzzling small area molecular structures discovered by 
\citet{hh02,hh04,hh06}.

To fully characterize the impact of the turbulence on the chemistry of the 
gas, it is important to determine the amount of mixing which occurs 
\citep{xal95,wla02}. Our current simulations have not allowed us to do this, 
since they are strictly Eulerian in nature, but we plan to address this point in 
future work. 

We have also examined the sensitivity of our results to the choice of several
of our main input parameters: the initial temperature of the gas, $T_{i}$,
the box size $L$, the initial magnetic field strength $B_{\rm i}$, 
the initial turbulent velocity dispersion $v_{\rm rms, i}$, and the initial 
density $n_{\rm i}$.

We find very little sensitivity to the initial gas temperature. The gas in our 
simulations cools rapidly and reaches thermal equilibrium within only 
$0.05 \: {\rm Myr}$ and at later times displays no memory of its initial 
temperature. Indeed, if we look solely at the mean gas temperature at
times $t > 0.05 \: {\rm Myr}$, then we find that it behaves rather like a 
polytrope, with $T \propto n^{\gamma_{\rm eff} - 1}$ and 
$\gamma_{\rm eff} \simeq 0.8$. However, at low gas densities, the scatter 
of the actual gas temperature around this mean value is considerable,
and the polytropic approximation does not capture the full range of 
behaviour of the gas.

Simulations performed with different values for the box size $L$ do show
some differences in behaviour, with $\mHt$ forming faster when $L$ is 
small. This appears to be due to the fact that altering $L$ alters the turbulent
crossing time, which in turn alters the time required for large density
enhancements to build up within the box. Nevertheless, in every case 
our main result -- that $\mHt$ forms rapidly, on a timescale of a few 
megayears -- remains the same.

Varying the strength of the initial magnetic field also leads to some 
differences in behaviour, although again in every case we find the
same basic result. $\mHt$ forms more rapidly and in greater
quantities when $B_{\rm i} = 0$ than when $B_{\rm i} \neq 0$, apparently 
because large overdensities develop faster in the former case than in the 
latter. Surprisingly, $\mHt$ also forms more rapidly when the magnetic
field is strong and the gas is subcritical than when it is weak (but still 
dynamically significant) and the gas is supercritical. This appears to be
due to the fact that when the field is strong, it remains highly ordered, 
causing gas to flow preferentially along the field lines and leading to
the formation of dense shocked layers that are oriented perpendicularly 
to the mean field. An effect of this kind was previously noted by \citet{hmk01} 
in their simulations of isothermal MHD turbulence.

Decreasing the strength of the turbulence by decreasing $v_{\rm rms, i}$ 
lengthens the $\mHt$ formation timescale, as it again takes longer to build 
up the same amount of dense structure. However, $\mHt$ formation remains 
rapid compared to quiescent models (cf.\ paper I) and mass-weighted mean 
molecular fractions of $\sim 40\%$ can be produced in less than $4 \: {\rm Myr}$ 
as long as $v_{\rm rms, i} \simgreat 2.5 \: {\rm km} \: {\rm s^{-1}}$. Since observations 
of molecular clouds on scales $\sim 20 \: {\rm pc}$ find one-dimension velocity 
dispersions of $2$--$7 \: {\rm km} \: {\rm s^{-1}}$ \citep{srby87},
corresponding to values of $v_{\rm rms} \simeq 3.5$--$12 \: {\rm km} \: 
{\rm s^{-1}}$, this implies that $\mHt$ formation will be rapid in real clouds.

The input parameter to which our results are most sensitive is the initial number 
density $n_{\rm i}$. Decreasing $n_{\rm i}$ substantially decreases the amount 
of dense gas produced in the simulations, which has the effect of dramatically 
reducing the production of $\mHt$.  Based purely on our results, we would expect 
significant $\mHt$ formation to occur only in clouds with mean densities greater than 
$\sim 10 \: {\rm cm^{-3}}$, although we caution that this is no better than an
order of magnitude estimate at this point.

In summary, we suggest that the key to understanding the formation of molecular 
clouds is understanding the process (or processes) that produce large, dense clouds of 
turbulent {\em atomic} gas, since our results demonstrate that such clouds will very quickly 
become molecular. Our results also show that the many transient overdense regions that 
are created by the turbulence play a central role in bringing about this rapid rate of 
$\mHt$ formation.

\acknowledgments
The authors would like to acknowledge valuable discussions on various aspects of this
work with J.\ Black, R.\ Garrod, R.\ Klessen, C.\ Lintott, H.\ Liszt, J.\ Niemeyer, A.\ Rosen, 
W.\ Schmidt and M.\ Smith. Comments by F.\ Shu, R.\ Klein, and others at a meeting of
the Berkeley-Santa Cruz-Ames Star Formation Center first inspired this
work. The simulations discussed  
in this paper were performed on the Parallel Computing Facility of the AMNH and on an
Ultrasparc III cluster generously donated to the AMNH by Sun Microsystems.  We would 
like to thank T.\ Grant, D.\ Harris, S.\ Singh, and, in particular, J.\ Ouellette 
for their invaluable technical assistance at various points during the simulation runs. 
Financial support for this work was provided by NASA grant NAG5-13028 and NSF grant 
AST-0307793.

\begin{deluxetable}{ccccccc}
\tablecaption{Input parameters used for each simulation. 
\label{turb_runs}}
\tablewidth{0pt}
\tablehead{\colhead{Run}  & \colhead{$L$ (pc)}
& \colhead{$n_{\rm i}$ ($\rm{cm^{-3}}$)}  
& \colhead{$T_{\rm i}$ (K)} & \colhead{$B_{\rm i}$ ($\mu$G)} 
& \colhead{$v_{\rm rms, i}$ (km s$^-1$)} & \colhead{Notes}}
\startdata
MT64    & 20 & 100 & 1000 & 5.85  & 10.0 &  \\
MT128   & 20 & 100 & 1000 & 5.85  & 10.0 &  \\
MT256   & 20 & 100 & 1000 & 5.85  & 10.0 &  \\
MT512   & 20 & 100 & 1000 & 5.85  & 10.0 &  \\
MT64-ng     & 20 & 100 & 1000 & 5.85  & 10.0 & 1 \\
MT128-ng    & 20 & 100 & 1000 & 5.85  & 10.0 & 1 \\
MT256-ng    & 20 & 100 & 1000 & 5.85  & 10.0 & 1 \\
MT512-ng    & 20 & 100 & 1000 & 5.85  & 10.0 & 1 \\
MT256-nr    & 20 & 100 & 1000 & 5.85  & 10.0 & 2 \\
MT256-RT    & 20 & 100 & 1000 & 5.85  & 10.0 & 3 \\
MT256-th3e2 & 20 & 100 & 1000 & 5.85 & 10.0 & 4 \\
MT256-th1e3 & 20 & 100 & 1000 & 5.85 & 10.0 & 4 \\
MT256-th3e3 & 20 & 100 & 1000 & 5.85 & 10.0 & 4 \\
MT256-T100 & 20 & 100 & 100 & 5.85 & 10.0 & \\
MT256-L10 & 10 & 100 & 1000 & 5.85 & 10.0 & \\
MT256-L30 & 30 & 100 & 1000 & 5.85 & 10.0 & \\
MT256-L40 & 40 & 100 & 1000 & 5.85 & 10.0 & \\
HT256       & 20 & 100 & 1000 & 0.0   & 10.0 & \\
MT256-Bx2   & 20 & 100 & 1000 & 11.7  & 10.0 & \\ 
MT256-Bx10  & 20 & 100 & 1000 & 58.5  & 10.0 & \\
MT256-v1    & 20 & 100 & 1000 & 5.85  &  1.0 & \\
MT256-v2.5  & 20 & 100 & 1000 & 5.85  &  2.5 & \\
MT256-v5    & 20 & 100 & 1000 & 5.85  &  5.0 & \\
MT256-n10   & 20 &  10 & 1000 & 0.585  & 10.0 & \\
MT256-n30   & 20 &  30 & 1000 & 1.755  & 10.0 & \\
MT256-RT-n10 & 20 &  10 & 1000 & 0.585 & 10.0 & 3 \\
\enddata
\tablecomments{1: runs with self-gravity disabled; 2: run with $\chi = 0.0$;
3: runs using the six-ray shielding approximation;
4: runs with $\mHt$ formation switched off in gas with $n < n_{\rm th}$}
\end{deluxetable}

\begin{deluxetable}{lcccc}
\tablecaption{$t_{\rm res}$, $t_{\rm f}$, and associated values of $\Htmass$
for all runs in Table~\ref{turb_runs}. \label{xh2_at_end}}
\tablewidth{0pt}
\tablehead{\colhead{Run}  & 
\colhead{$t_{\rm res}$ (Myr)} &
\colhead{$\Htmass (t_{\rm res})$} &
\colhead{$t_{\rm f}$ (Myr)} &
\colhead{$\Htmass (t_{\rm f})$}}
\startdata
MT64        & 1.17 & 0.12 & 7.92 & 0.80 \\
MT128       & 0.98 & 0.15 & 7.92 & 0.40 \\
MT256       & 1.08 & 0.18 & 3.01 & 0.42 \\
MT512       & 1.49 & 0.30 & 2.54 & 0.42 \\
MT64-ng     & 0.35 & 0.04 & 7.93 & 0.36 \\
MT128-ng    & 1.77 & 0.27 & 7.92 & 0.30 \\
MT256-ng    & 1.78 & 0.30 & 7.92 & 0.23 \\
MT512-ng    & 1.81 & 0.35 & 2.85 & 0.41 \\
MT256-nr    & 0.95 & 0.18 & 3.48 & 0.61 \\
MT256-RT    & 0.98 & 0.19 & 2.85 & 0.55 \\
MT256-th3e2 & 1.08 & 0.15 & 4.66 & 0.35 \\
MT256-th1e3 & 1.08 & 0.11 & 3.23 & 0.28 \\
MT256-th3e3 & 1.08 & 0.05 & 2.06 & 0.12 \\
MT256-T100  & 1.08 & 0.18 & 2.06 & 0.35 \\
MT256-CR    & 1.08 & 0.19 & 2.96 & 0.43 \\
MT256-L10   & ---  & ---  & 3.96 & 0.18 \\
MT256-L30   & 0.92 & 0.11 & 3.26 & 0.46 \\
MT256-L40   & 0.66 & 0.05 & 3.01 & 0.41 \\
HT256       & 0.60 & 0.25 & 1.97 & 0.57 \\
MT256-Bx2   & 1.46 & 0.33 & 7.92 & 0.21 \\
MT256-Bx10  & ---  & ---  & 3.19 & 0.43 \\
MT256-v1    & 5.13 & 0.21 & 7.48 & 0.41 \\
MT256-v2.5  & 2.66 & 0.22 & 5.13 & 0.48 \\
MT256-v5    & 1.62 & 0.20 & 5.61 & 0.47 \\
MT256-n10   & --- & --- & 7.92 & $5.9 \times 10^{-4}$ \\
MT256-n30   & --- & --- & 3.01 & 0.1 \\
MT256-RT-n10 & --- & --- & 2.76 & 0.07 \\
\enddata
\tablecomments{$t_{\rm res}$ is the time at which the Truelove criterion
is first violated during the course of the run; when no value is given,
this indicates that the criterion was never violated. $t_{\rm f}$ is the
time at which the simulation was stopped.}
\end{deluxetable}

\begin{deluxetable}{lcc}
\tablecaption{$f_{\rm res}$ and $f_{\rm res, \mHt}$ for all runs in
Table~\ref{turb_runs}. \label{fres_at_end}}
\tablewidth{0pt}
\tablehead{\colhead{Run}  & 
\colhead{$f_{\rm res}(t_{\rm f})$} &
\colhead{$f_{\rm res, \mHt}(t_{\rm f})$}}
\startdata
MT64     & 0.640 & 0.558 \\
MT128    & 0.974 & 0.935 \\
MT256    & 0.996 & 0.990 \\
MT512    & 0.998 & 0.994 \\
MT64-ng  & 0.999 & 1.000 \\
MT128-ng & 1.000 & 1.000 \\
MT256-ng & 1.000 & 1.000 \\
MT512-ng & 1.000 & 1.000 \\
MT256-nr & 0.883 & 0.877 \\
MT256-RT & 0.996 & 0.992 \\
MT256-th3e2 & 0.998 & 0.993 \\
MT256-th1e3 & 0.997 & 0.988 \\
MT256-th3e3 & 0.999 & 0.992 \\
MT256-T100  & 0.997 & 0.992 \\
MT256-CR  & 0.997 & 0.993 \\
MT256-L10 & 1.000 & 1.000 \\
MT256-L30 & 0.987 & 0.971 \\
MT256-L40 & 0.985 & 0.963 \\
HT256 & 0.992 & 0.986 \\
MT256-Bx2 & 1.000 & 1.000 \\
MT256-Bx10 & 1.000 & 1.000 \\
MT256-v1 & 0.936 & 0.847 \\
MT256-v2.5 & 0.968 & 0.931 \\
MT256-v5 & 0.979 & 0.953 \\
MT256-n10 & 1.000 & 1.000 \\
MT256-n30 & 1.000 & 1.000 \\
MT256-nRT-n10 & 1.000 & 1.000 \\
\enddata
\tablecomments{$f_{\rm res}$ is the fraction of the total gas mass
that is situated in zones that satisfy the Truelove criterion, 
computed at the end of each simulation; $f_{\rm res, \mHt}$ is
the fraction of the total $\mHt$ mass situated in these zones}
\end{deluxetable}

\end{document}